\documentclass[pra,onecolumn,nofootinbib,superscriptaddress]{revtex4-2}
\usepackage{calrsfs}
\usepackage{ulem}
\usepackage{graphics}
\usepackage{epstopdf}
\usepackage{bm}
\usepackage{bbm}
\usepackage{amssymb}
\usepackage{epstopdf}
\usepackage{graphicx}
\usepackage[caption=false]{subfig}
\usepackage{amsmath}
\usepackage{color}
\usepackage{array}

\begin{document}

\title{Quantitative phase gradient microscopy with spatially entangled photons}

\author{Yingwen \surname{Zhang}}
\email{yzhang6@uottawa.ca}
\affiliation{Nexus for Quantum Technologies, University of Ottawa, Ottawa ON, Canada, K1N6N5}
\affiliation{National Research Council of Canada, 100 Sussex Drive, Ottawa ON, Canada, K1A0R6}
\affiliation{Joint Centre for Extreme Photonics, National Research Council and University of Ottawa, Ottawa, Ontario, Canada}

\author{Paul-Antoine \surname{Moreau}}
\email{pa.moreau@gs.ncku.edu.tw}
\affiliation{Department of Physics, National Cheng Kung University, Tainan, 70101, Taiwan}
\affiliation{Center for Quantum Frontiers of Research and Technology, NCKU, Tainan, 70101, Taiwan}

\author{Duncan \surname{England}}
\affiliation{National Research Council of Canada, 100 Sussex Drive, Ottawa ON, Canada, K1A0R6}

\author{Ebrahim \surname{Karimi}}
\affiliation{Nexus for Quantum Technologies, University of Ottawa, Ottawa ON, Canada, K1N6N5}
\affiliation{National Research Council of Canada, 100 Sussex Drive, Ottawa ON, Canada, K1A0R6}
\affiliation{Joint Centre for Extreme Photonics, National Research Council and University of Ottawa, Ottawa, Ontario, Canada}
\affiliation{Institute for Quantum Studies, Chapman University, Orange, California 92866, USA}

\author{Benjamin \surname{Sussman}}
\affiliation{National Research Council of Canada, 100 Sussex Drive, Ottawa ON, Canada, K1A0R6}
\affiliation{Nexus for Quantum Technologies, University of Ottawa, Ottawa ON, Canada, K1N6N5}
\affiliation{Joint Centre for Extreme Photonics, National Research Council and University of Ottawa, Ottawa, Ontario, Canada}

\begin{abstract} 
We present an entanglement-based quantitative phase gradient microscopy technique that employs principles from quantum ghost imaging and ghost diffraction. In this method, a transparent sample is illuminated by both photons of an entangled pair--one detected in the near-field (position) and the other in the far-field (momentum). Due to the strong correlations offered by position–momentum entanglement, both conjugate observables can be inferred nonlocally, effectively enabling simultaneous access to the sample’s transmission and phase gradient information. This dual-domain measurement allows for the quantitative recovery of the full amplitude and phase profile of the sample. Unlike conventional classical and quantum phase imaging methods, our approach requires no interferometry, spatial scanning, microlens arrays, or iterative phase-retrieval algorithms, thereby circumventing many of their associated limitations. Furthermore, intrinsic temporal correlations between entangled photons provide robustness against dynamic and structured background light. We demonstrate quantitative phase and amplitude imaging with a spatial resolution of 2.76\,$\mu$m and a phase sensitivity of $\lambda/100$ using femtowatts of illuminating power, representing the highest performance reported to date in quantum phase imaging. This technique opens new possibilities for non-invasive imaging of photosensitive samples, wavefront sensing in adaptive optics, and imaging under complex lighting environments.
\end{abstract}
\maketitle

\section{Introduction}

Phase contrast microscopy, first developed by Frits Zernike in the 1930s\cite{Zernike1,Zernike2}, was a groundbreaking technique that enabled the visualization of transparent specimens by converting subtle variations in optical path length within a transparent sample into intensity variations in an image. This method revolutionized biological imaging by producing high-contrast images of live cells and tissues without the need for staining. While phase contrast microscopy provides qualitative information, it does not yield absolute measurements of optical phase. Building on this foundation, quantitative phase imaging (QPI) has emerged as a powerful class of techniques that measure optical path length variations with nanometric sensitivity and spatial precision and has found widespread application in biomedical imaging and materials science. The primary approaches to achieving QPI are through interferometry, wavefront sensing, and phase-retrieval algorithms~\cite{Nguyen2022,Chaumet2024}. Across these approaches, there is often a trade-off between spatial resolution, phase sensitivity, acquisition speed, hardware complexity, and robustness. For instance, interferometric methods~\cite{Huang2024} offer high phase accuracy but are susceptible to environmental noise and instability in the reference beam. Wavefront sensing techniques such as the Shack-Hartmann sensor~\cite{Shack1971,Gong2017} operate without a reference beam and have relatively simple setups, yet their spatial resolution is limited by the microlens array. Phase retrieval algorithms can be compact and cost-effective but often have slow imaging speeds due to requiring multiple intensity measurements—at different planes (e.g. transport of intensity equation~\cite{Streibl1984} and Gerchberg–Saxton algorithm~\cite{Gerchberg1972,Fienup1982}) or at different angles (e.g. Fourier ptychography~\cite{Zheng2013,Tian2015} and differential phase contrast microscopy~\cite{Tian2015_2,Mehta2009})—and the algorithms may sometimes suffer from convergence instability.

Utilizing the properties of quantum entangled photons for sensing and imaging applications has been an active area of research in recent decades~\cite{Genovese_2016,Berchera_2019,Moreau2019,Degen2017,Pirandola2018}. Quantum light offers a range of potential advantages over its classical counterpart, this includes super-resolution~\cite{Tsang2009,Shin2011,Rozema2014,Unternahrer2018,Tenne2019}, robustness against noise~\cite{Brida2010,Samantaray2017,Zhang2020,Defienne2021_2,Zhao2022}, allows probing and imaging a sample at different wavelengths~\cite{Aspden2015,Ryan2024} and technical advantages such as immunity to dispersion in optical coherence tomography~\cite{Nasr2003,Yepiz2022}, achieving higher spatial and spectral resolution in hyperspectral imaging~\cite{Zhang20222} and larger volumetric depth in 3D imaging~\cite{Zhang2022,Zhang2024}.

The use of entangled photons has also been extended to phase imaging. Interferometric quantum phase retrieval techniques include the use of photonic N00N states~\cite{Ono2013,Israe2014} and the interference between successive laser passes of spontaneous parametric down-conversion (SPDC) events~\cite{Black2023}. Phase imaging has also been demonstrated by holography with entangled photons~\cite{Defienne2021,Guillaume2023}. Non-interferometric methods using entangled photons include ghost diffraction~\cite{Abouraddy2004}, single-pixel ghost imaging~\cite{Sephton2023}, using transport of intensity equation~\cite{Chien2015,Ortolano2023}, Fourier ptychography~\cite{Aidukas2019}, and asymmetric illumination~\cite{Hodgson2022}. These techniques offer various advantages such as sub-shot-noise sensitivity~\cite{Ono2013,Israe2014,Ortolano2023,Chien2015}, background-noise tolerance~\cite{Aidukas2019,Defienne2021}, resilience to global phase drifts~\cite{Guillaume2023} or allowing for scanning-free measurement~\cite{Hodgson2022}. Recently, measurement of the biphoton spatial wavefunction of SPDC has also been realized through using a Shack-Hartmann sensor~\cite{Zheng2024} and the Gerchberg–Saxton algorithm~\cite{Dehghan2024}. Despite many potential advantages, many of these quantum techniques still suffer from limitations similar to those of their classical analogs, such as sensitivity to interference instabilities or the need for taking multiple sample images. A demonstration of quantum QPI achieving spatial resolution and phase sensitivity comparable to that of classical QPI microscopes also remains lacking.

In this work, we present a proof-of-concept demonstration of an entanglement-based quantitative phase gradient microscopy technique that harnesses the principles of quantum ghost imaging and ghost diffraction. Our method exploits position–momentum entanglement in photon pairs generated via SPDC, which are strongly correlated in position and anti-correlated in momentum. The position of the signal photon is measured in the near-field (NF) of the sample, while the momentum of the idler photon is measured in the far-field (FF). Although only one observable is recorded from each photon, their joint correlations provide simultaneous nonlocal access to both position and momentum information. This approach effectively unifies ghost imaging (recovering idler position via signal detection) and ghost diffraction (recovering signal momentum via idler detection). By leveraging the correspondence between phase gradients in the NF and momentum shifts in the FF, we can quantitatively reconstruct the phase profile of the sample. This technique eliminates the need for interferometry, spatial scanning, microlens arrays, or iterative phase retrieval algorithms, thus avoiding many of the associated limitations of many conventional QPI techniques. We refer to this method as quantum correlation phase gradient microscopy (QCPGM).

Using QCPGM, we demonstrate quantitative phase microscopy with a spatial resolution of 2.76\,$\mu$m and phase sensitivity of $\lambda/100$ using just 100\,fW of illuminating power. To the best of our knowledge, this is the highest performance reported to date in quantum phase imaging. This capability is exemplified by the acquisition of both phase and amplitude images of epithelial cheek cells. The technique is also shown to be resilient to complex background illumination conditions. The broad applicability of QCPGM includes phase imaging of photosensitive samples, wavefront sensing for adaptive optics and imaging under complex lighting environments.

\section{Experimental Results}
\subsection{Experimental Concept}
\begin{figure}
    \centering
    \includegraphics[width=1\linewidth]{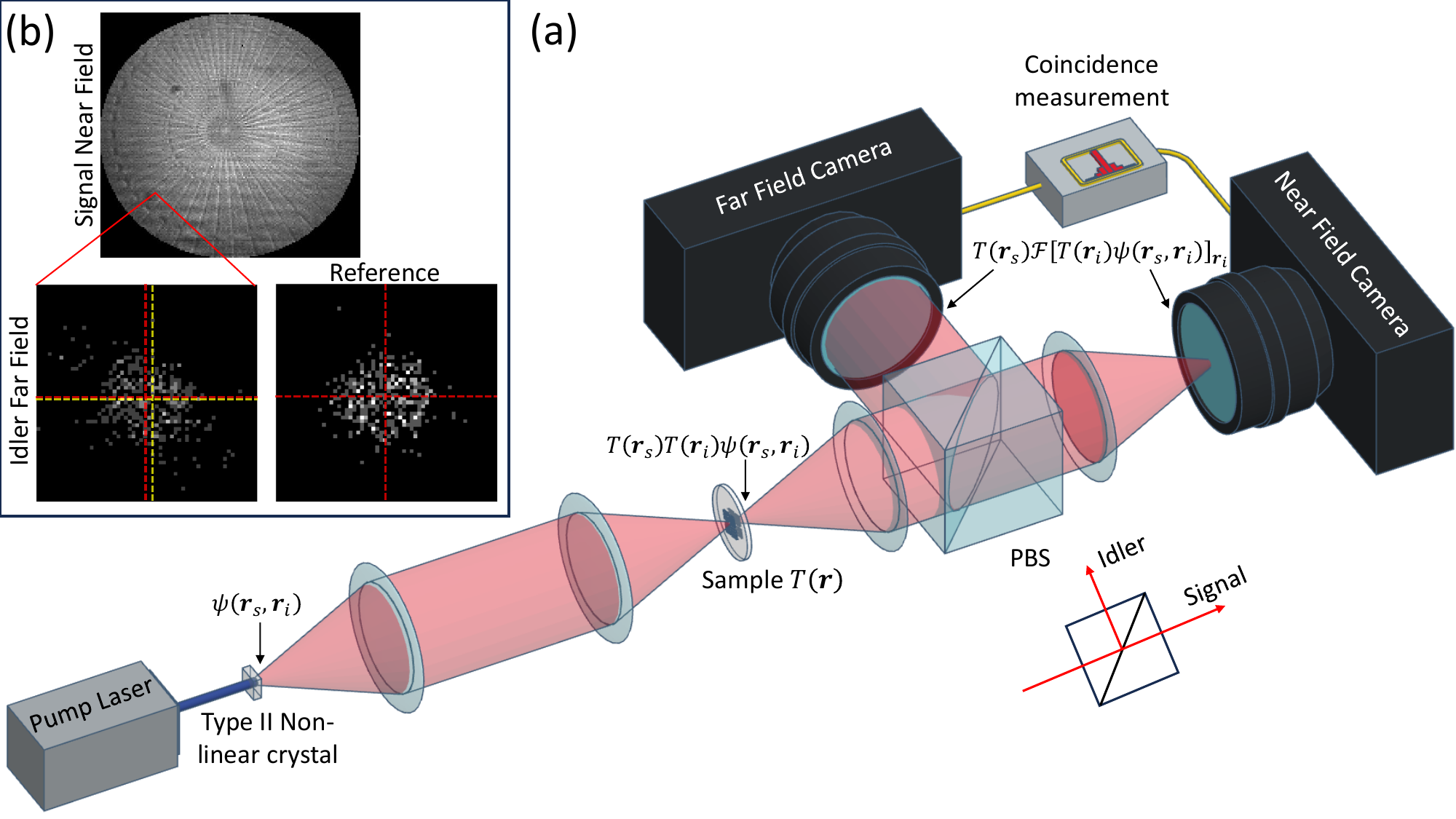}
    \caption{\textbf{Conceptual setup of QCPGM} (a) Spatially entangled photon pairs with orthogonal polarization are generated through a Type II nonlinear crystal. After illuminating a sample, the photon pairs are separated with a polarizing beam splitter (PBS) such that the NF and FF of the target can be imaged separately by the signal and idler photons, respectively. A time correlation measurement is performed to identify photon pairs between all photons captured between the two cameras. (b)  Brightfield image of a phase target captured by the NF camera and the image formed by all idler photons in the FF that are detected in coincidence with signal photons that passed through the indicated NF pixel. The FF beam centroid is indicated by the yellow crosshair and is shown relative to the centroid of the reference, captured for the same NF pixel with no phase target in place, indicated by the red crosshair. The change in centroid is proportional to the phase gradient at the indicated NF pixel.}
    \label{Concept}
\end{figure}

The conceptual setup of QCPGM is depicted in Fig.~\ref{Concept}. Position-momentum entangled photon pairs, denoted as ``signal" and ``idler" photons, with orthogonal polarization are generated through Type-II SPDC. In the low gain regime, the position-momentum entangled state of SPDC in position space $\bm{r} = (x,y)$ can be written as:
\begin{equation}
    |\Psi\rangle = \int\int \psi(\bm{r}_s,\bm{r}_i) |\bm{r}_s,\bm{r}_i\rangle\, d^2r_s d^2r_i,
\end{equation}
where $\psi(\bm{r}_s,\bm{r}_i)$, the biphoton correlation function, under the double-Gaussian approximation is \cite{Law2004,Chan2007,Schneeloch2016}:
\begin{equation}
    \psi({\bm{r}}_s,{\bm{r}}_i) \sim \exp\left(\frac{-|\bm{r}_s-\bm{r}_i|^2}{2\delta_r^2}\right)\exp\left(-2\delta_k^2|\bm{r}_s+\bm{r}_i|^2\right),
\end{equation}
with $\delta_k \approx 1/(2\sigma_p)$, $\sigma_p$ being the pump beam width, and $\delta_r\approx\sqrt{\frac{2\alpha L \lambda_p}{\pi}}$ where $L$ is the crystal length, $\lambda_p$ is the pump wavelength, and $\alpha=0.455$ is a constant factor from the Gaussian approximation of the phase matching function in momentum space.

The photon pairs jointly illuminate a target with complex transmission $T(\bm{r}) = A(\bm{r})e^{i\phi(\bm{r})}$, placed in the NF of the nonlinear crystal. The coincidence pattern directly after the target is given by
\begin{equation}
    C(\bm{r}_s,\bm{r}_i) \propto  \Big|T(\bm{r}_s) T(\bm{r}_i)\psi({\bm{r}}_s,{\bm{r}}_i) \Big|^2.
\end{equation}

A polarizing beamsplitter then separates the photons, sending the signal photon through a 4f system to a time-tagging camera (TPX3CAM~\cite{Nomerotski2019,ASI}) for NF imaging and the idler photon for FF imaging on another camera. Knowing each photon pair is created at the same instance in time, they are identified through a time-correlation measurement using the detection time information from the camera.

The resulting coincidence pattern for each combination of signal photon position $\bm{r}_s$ and idler photon momentum $\bm{k}_i=(u_i,v_i)$ is given by
\begin{equation}
    C(\bm{r}_s,\bm{k}_i) \propto  \Big|T(\bm{r}_s) \mathcal{F} \left[T(\bm{r}_i)\psi({\bm{r}}_s,{\bm{r}}_i) \right]_{\bm{r}_i} \Big|^2,
    \label{quantumI}
\end{equation}
where $\mathcal{F}\left[f(\bm{r}_s,\bm{r}_i)\right]_{\bm{r}_i}$ is the 2-dimensional Fourier transform with respect to ${\bm{r}_i}$. 

The expectation value of the phase gradient $\nabla[\phi(\bm{r}) + \phi_0(\bm{r})]$ is proportional to the beam centroid in the FF, which in the paraxial limit is
\begin{equation}
\begin{pmatrix}
\langle \frac{\partial\phi(\bm{r}_s)}{\partial x} \rangle\\
\langle \frac{\partial\phi(\bm{r}_s)}{\partial y} \rangle
\end{pmatrix}
= \frac{1}{f}
\begin{pmatrix}
\mathcal{U}(\bm{r}_s) - \mathcal{U}_0(\bm{r}_s)\\
\mathcal{V}(\bm{r}_s) - \mathcal{V}_0(\bm{r}_s)
\end{pmatrix},
\end{equation}
with $f$ the lens focal length, $\mathcal{U}(\bm{r}_s)$ and $\mathcal{V}(\bm{r}_s)$ the centroid positions in the $\hat{u}$ and $\hat{v}$ directions for all idler photons that are correlated with the signal photons detected at position $\bm{r}_s$ and $\mathcal{U}_0(\bm{r})$ and $\mathcal{V}_0(\bm{r})$ being the reference centroid positions obtained without the target in place.


Finally, the phase $\phi(\bm{r})$ is reconstructed by solving the two-dimensional partial differential equations $\frac{\partial\phi(\bm{r})}{\partial x}$ and $\frac{\partial\phi(\bm{r})}{\partial y}$. For this, we use the Frankot and Chellappa method~\cite{Frankot1988,Gong2017}
\begin{equation}
\phi(\bm{r}) = \mathcal{F}^{-1}\left[\frac{u \mathcal{F}\left[\frac{\partial\phi(\bm{r})}{\partial x} \right] + v \mathcal{F}\left[\frac{\partial\phi(\bm{r})}{\partial y} \right]}{i(u^2+v^2)} \right],
\end{equation}
where $\mathcal{F}[\cdot]$ and $\mathcal{F}^{-1}[\cdot]$ represent the Fourier and inverse Fourier transform, respectively. Compared to the finite difference method, the Frankot and Chellappa method provides better handling of real-world data, which can contain noisy, non-integrable gradient fields, and, based on the Fast-Fourier transform, it is more computationally efficient. A derivation of this method can be found in the Supplementary. 



\subsection{Quantitative phase measurement}
\begin{figure}
    \centering
    \includegraphics[width=1\linewidth]{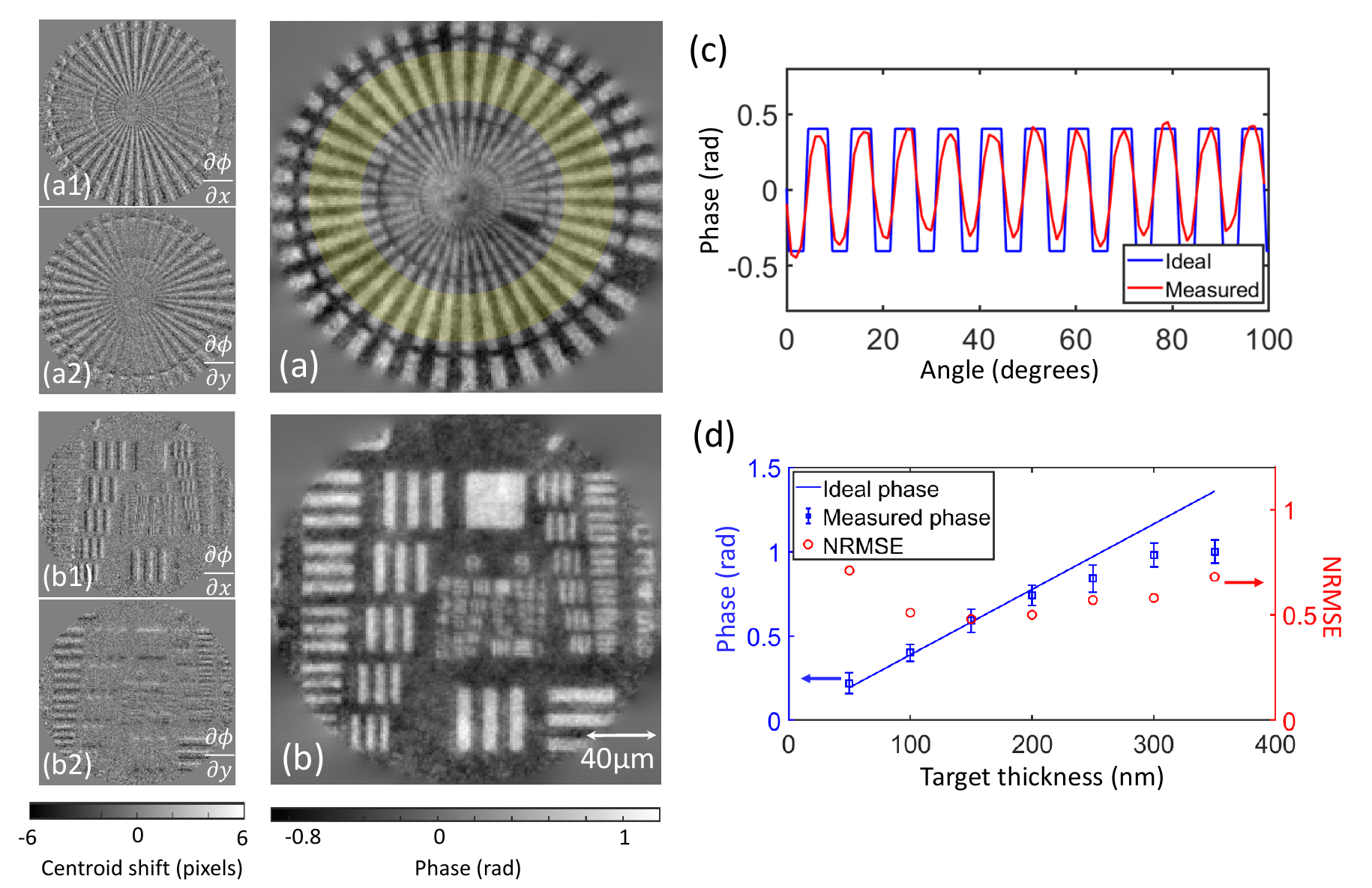}
    \caption{\textbf{Results from imaging resolution phase targets} (a, b) Recovered phase images for a Star (a) and 1951 USAF (b) resolution phase target with 200\,nm feature height. The corresponding measured phase gradient in the horizontal and vertical direction used to recover the phase image are shown in (a1, b1) and (a2, b2), respectively. (c) Cross-section from 0 to 100 degrees of the yellow highlighted regions of (a) compared to an ideal phase measurement of this target. (d) Measured phase (blue square) and NRMSE (red circle) as a function of the target thickness for the Star target. The phase is measured as the mean difference between the maxima and minima in the cross-section of the yellow highlighted regions, as indicated in (a), and NRMSE compares the similarity between the measured phase profile and that of the ideal profile, as shown in (c). The bright-field image of the Star phase target is shown in Fig.\,\ref{Fig4}, and the full cross-sections are presented in the Supplementary material. The measured phase of the USAF target can also be found in the Supplementary. Data acquisition time for all images shown in this and subsequent figures is 500\,s. Background contribution has been corrected for all results that are shown.} 
    \label{Fig2}
\end{figure}

To validate the accuracy of the phase recovery process we used quantitative phase targets (from Benchmark technologies\,\cite{Benchmark}) with transparent polymer features, of refractive index 1.50 at 810\,nm, ranging in height from heights ranging from 50\,nm to 350\,nm, with two different patterns, the star and 1951 USAF. Having the target illuminated at a power of $\sim100$\,fW, we find the method is accurate from a target thickness of 50\,nm, or $\sim\lambda$/30 at 810\,nm (the phase difference between 50\,nm of polymer and air) up to 250\,nm, or $\lambda$/6.5. The smallest spatial feature observed in the microscope is group 8-4 of the 1951 USAF target, which corresponds to resolving line pairs separated by 2.76\,$\mu$m. The results for this are shown in Fig.~\ref{Fig2}. Based on the uncertainties in the phase measurement of approximately 0.06\,rad, we would expect the technique to still be phase sensitive at approximately $\lambda$/100. The uncertainties in the phase measurement can be reduced by increasing the data acquisition time or using a higher-efficiency camera, which will enable even better phase accuracy and sensitivity. A more detailed discussion on this is given in the Discussions section. Note that we are not yet imaging at the diffraction limit of the imaging system, which, given the numerical aperture of the lens arrangement, is expected to be $\sim 1$\,$\mu$m. This is a result of the large pixel size of the camera ($55$\,$\mu$m) limiting the achievable resolution. 

As a measure of the similarity between the recovered phase profile and that of an ideal phase profile, which has a constant phase across the target features, we determine the normalized root mean squared error (NRMSE) between the two phase profiles, which is defined as
\begin{equation}
    \text{NRMSE} = \frac{\sqrt{\frac{1}{N}\sum_{j=1}^N (O_j - E_j)^2}}{\bar{E}},
\end{equation}
where $O_j$ and $E_j$ are the observed and expected values, respectively. $\bar{E}$ is the mean of the expected values. 

The NRMSE for the Star targets of different heights are shown as red circles in Fig.~\ref{Fig2}(d), we see that the NRMSE deteriorates for both smaller and larger feature heights. This outcome aligns with expectations. Smaller feature heights amplify the influence of shot noise in relation to the subtle phase profile. For larger feature heights, discrepancies between the measured and ideal phase profiles can be attributed to several reasons. One is that the large phase jump will cause a very large phase gradient that diffracts photons to outside the numerical aperture of the FF imaging system, resulting in measuring a smaller centroid shift. The other being that the linear relationship between the phase gradient and the FF centroid shift no longer holds for larger diffraction angles, as the paraxial approximation is no longer valid. Therefore, the method is most accurate for targets that do not contain phase jumps of over $\sim\lambda/6$. It should be noted here that our analysis did not account for manufacturing irregularities in the quantitative phase targets, which may exhibit deviations of up to 10\% from the specified values, as stipulated by the manufacturer. 

\begin{figure}
    \centering
    \includegraphics[width=1\linewidth]{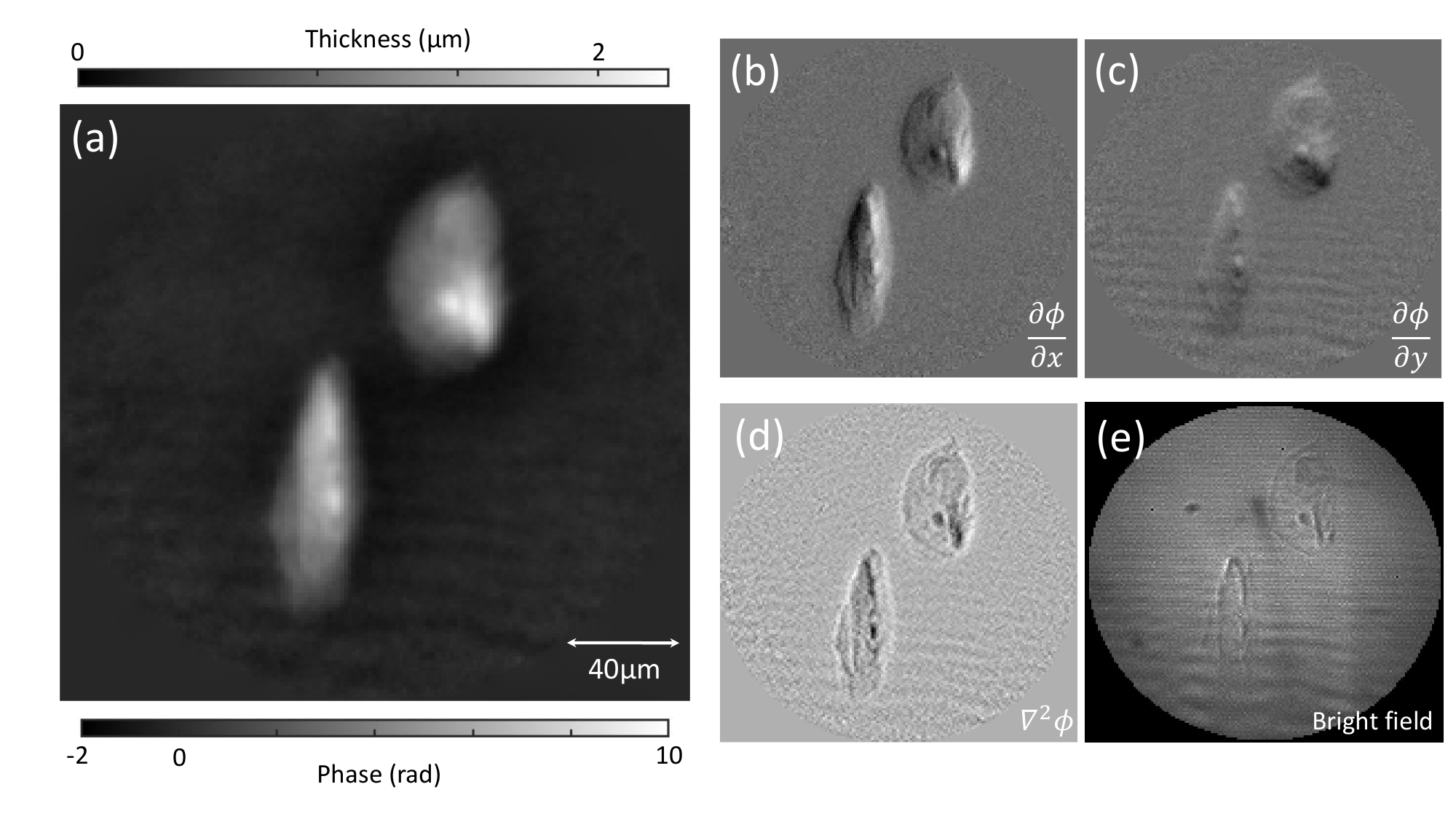}
    \caption{\textbf{Results of imaging cheek epithelial cells} (a) Phase image of cheek epithelial cells. (b, c) The measured phase gradient in the horizontal and vertical direction for recovering the phase image (a). (d) The Laplacian of the phase. (e) Image of the cells captured directly by the camera without performing correlation analysis, as if captured through a conventional bright-field microscope.}
    \label{Fig3}
\end{figure}

Lastly, in Fig.~\ref{Fig3} we show the phase imaging of cheek epithelial cells where a refractive index of 1.35~\cite{Gul2021} was assumed for the cells when estimating the cell thickness. 

\subsection{Dynamic background mitigation}
\begin{figure}
    \centering
    \includegraphics[width=1\linewidth]{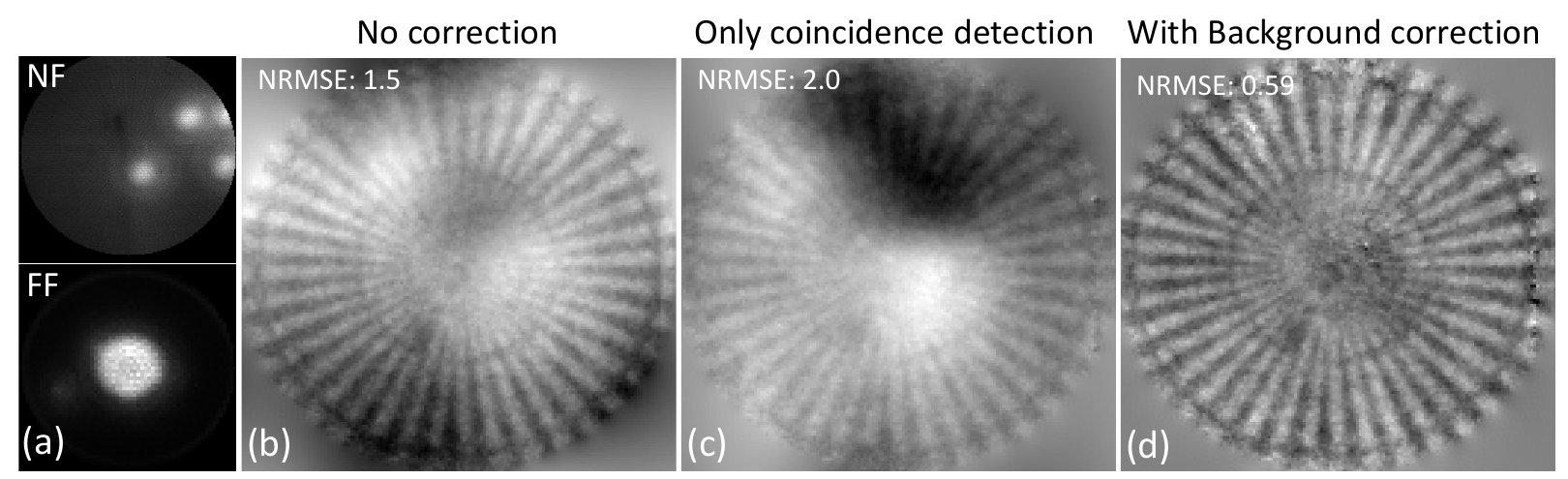}
    \caption{\textbf{Results demonstrating dynamic background mitigation} (a) NF and FF images obtained through direct imaging of the phase target with 200\,nm feature heights where a dynamic background light is added through an attenuated laser beam. (b) Expected phase image recovered with no background correction through a classical phase gradient microscope. (c) Recovered phase image with only coincidence detection. (d) Recovered phase images with full background correction. The cross-section for all phase images can be found in the supplementary materials.}
    \label{Fig4}
\end{figure}

Quantitative phase imaging techniques are highly susceptible to the influence of background light, which can obscure or distort the phase information of interest. Addressing the background typically involves either separately measuring the background light or applying Fourier filtering techniques, which require some prior knowledge of the background characteristics. These methods are often limited in their effectiveness, especially against dynamic backgrounds which can fluctuate with space and time. 

To emulate such a scenario, we introduce an attenuated laser beam into the setup, containing approximately 67\% of the photon flux of the SPDC, giving a signal to background ratio (SBR) of 1.49. The laser beam is repositioned over time to simulate a dynamic background; such a background could be introduced by fluorescence or auto-fluorescence in biological samples. The resultant brightfield image of a test phase target with 200\,nm feature heights, captured by the NF and FF camera, is shown in Fig.~\ref{Fig4}(a). The recovered phase image, corresponding to that expected from a classical phase gradient microscope (such as a Shack–Hartmann sensor) at equivalent spatial resolution to this QCPGM demonstration, is presented in Fig.\ref{Fig4}(b). In the presence of background light, the reconstructed phase is severely distorted, yielding a normalized root mean square error (NRMSE) of 1.5 when compared to the expected phase profile.

Quantum imaging techniques based on photon-pair correlations offer inherent resilience to background light, s coincidence detection selectively includes only photon pairs arriving within a defined temporal window~\cite{England2019,Zhao2022}. However, this advantage can diminish at high photon fluxes. Specifically, because accidental (uncorrelated) coincidences scale quadratically with total photon flux, the SBR achievable via coincidence detection may fall below that of direct photon counting under high-brightness conditions (see Supplementary Materials). In our experiment, with a pair generation rate of $\sim 3.5\times10^7$ pairs per second and a coincidence window of 20\,ns, the post-coincidence SBR is reduced to just 0.51. As a result, the reconstructed phase image becomes significantly distorted, with the NRMSE increasing to 2.0, as shown in Fig.~\ref{Fig4}(c).

Despite this, coincidence detection offers a unique advantage: the ability to estimate and subtract the background contribution directly from the data. As detailed in the Methods, this is achieved by measuring accidental coincidences using a temporal window offset from the true coincidence peak—in this case, 50\,ns earlier—where no true photon pairs are expected. The resulting background coincidence estimate can then be subtracted from the measured FF centroid to correct for background-induced artifacts. This procedure yields a significantly improved phase reconstruction, with an NRMSE of 0.59 (Fig.~\ref{Fig4}(d)). For reference, the NRMSE in the ideal case with no background laser is 0.50.

\section{Discussion}
In summary, we have demonstrated a scanning-free, non-interferometric quantitative phase microscopy technique that exploits the inherent position–momentum correlations of entangled photon pairs. By simultaneously acquiring the position and momentum information of the photon pairs through the principles of quantum ghost imaging and ghost diffraction, we recover the full amplitude and phase profile of a sample. This approach also offers intrinsic robustness to complex background lighting conditions.

Our proof-of-concept implementation achieved a spatial resolution of 2.76\,$\mu$m (corresponding to 362 line pairs per mm) and a phase sensitivity approaching $\lambda/100$, with quantitative phase accuracy better than $\lambda/30$ at a wavelength of 810\,nm, all under an illumination power of just 100\,fW. These results represent the highest-resolution and most phase-sensitive demonstrations to date in quantum phase imaging. Beyond imaging of photosensitive samples, the capabilities of QCPGM naturally lend themselves to broader applications, including high-resolution wavefront sensing for adaptive optics, especially in low-light or noise-sensitive environments.

QCPGM also does not suffer from many of the limitations inherent to other non-interferometric-based, classical or quantum, QPI techniques. It does not require multiple images of the sample to be taken, is not reliant on iterative algorithms which may not be convergent, and does not suffer from the resolution and sensitivity constraints in a Shack-Hartmann  wavefront sensor due to the use of microlens array (see the Supplementary for a theoretical comparison with the Shack-Hartmann sensor).


The ability to simultaneously access position and momentum information using entangled photons has only recently become feasible with the emergence of time-tagging single photon cameras. Much remains to be explored in this emerging direction, including the possibility of super-resolution imaging through ptychographic reconstruction of the momentum-space information, and volumetric phase imaging by combining QCPGM with the volumetric amplitude imaging technique as demonstrated in~\cite{Zhang2022,Zhang2024}.

The data acquisition speed of our technique is currently technically limited by the available camera technology. Our camera system exhibits a quantum efficiency of approximately 7\%, an 8\,ns time-resolution, and a maximum photon detection rate in the order of $10^7$ photons per second, as detailed in~\cite{Vidyapin2022}. A ten-fold improvement in each of the three camera parameters mentioned could potentially reduce the required data acquisition time to under a second. With the rapid advancements in single-photon detection technology, especially in superconducting nanowire cameras~\cite{Wollman2019,Oripov2023}, the prospect of camera technologies meeting these specifications is expected to be within reach in the coming years.

\section{Method}

\subsection{Experimental Setup}
\begin{figure}
    \centering
    \includegraphics[width=1\linewidth]{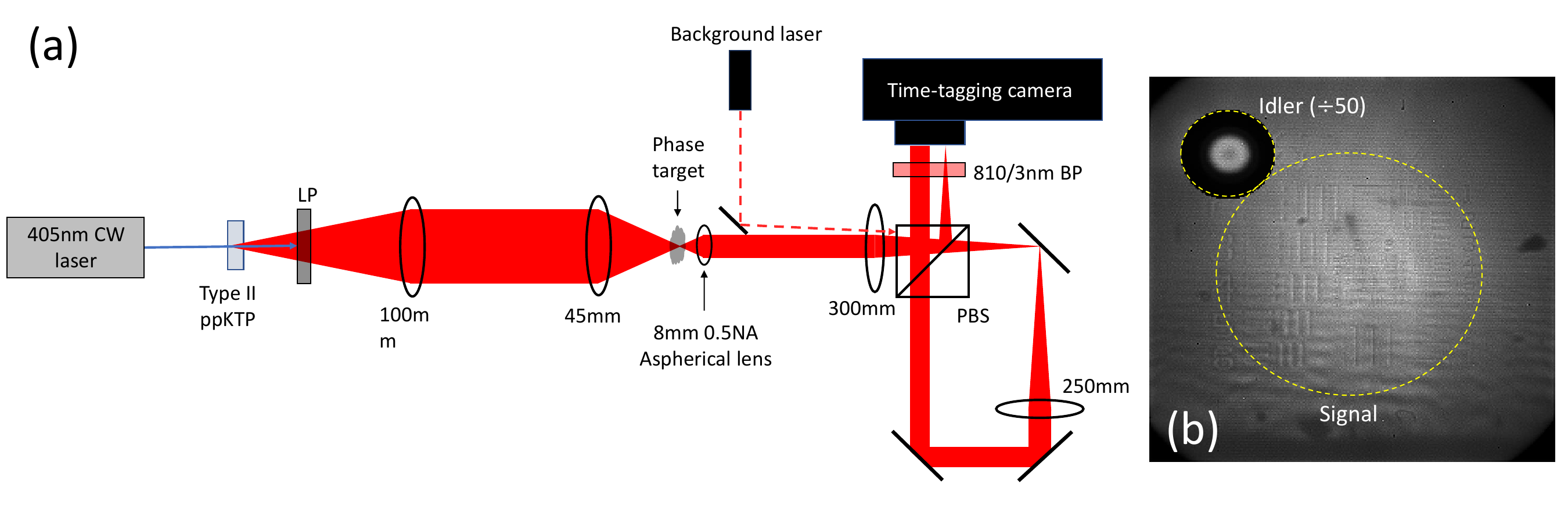}
    \caption{\textbf{Experimental setup of QCPGM} (a) PBS: polarizing beamsplitter, LP filter: longpass filter, BP filter: bandpass filter, ppKTP: periodically-poled potassium titanyl phosphate crystal. (b) Brightfield image of a phase target with 200\,nm feature height captured by the time-tagging camera. A time correlation measurement is then performed to identify photon pairs between all photons captured within the two highlighted circular regions. For viewing clarity, the idler beam shown here is scaled to be 50 times dimmer than its actual measurement, allowing the signal beam to be visible in the same image.}
    \label{Setup}
\end{figure}
The experimental setup is illustrated in Fig.~\ref{Setup}. A 20\,mW, 405\,nm continuous-wave (CW) laser, with a 1\,mm beam diameter, is used to pump a 1\,mm thick Type~II periodically-poled potassium titanyl phosphate (ppKTP) crystal to generate, through the process of SPDC, orthogonally polarized photon pairs at 810\,nm that are correlated in time and entangled in the position-momentum degrees of freedom. The photon pairs are directed through a 4f-imaging system to illuminate a phase target placed in the NF plane of the nonlinear crystal. The photons are then separated using a polarizing beamsplitter. The NF of the phase target is imaged onto a time-tagging camera (TPX3CAM) through the signal photons, while the FF of the target is projected onto a corner of the camera through a separate path using the idler photons. 

To identify photon pairs, time correlation measurement is conducted with a coincidence window of 20\,ns. Subsequently, a phase gradient measurement, as explained in the Experimental Concept section of the main text, is performed on each NF pixel to reconstruct the phase. Before conducting these measurements with the phase target, a one-time reference measurement without the phase target in place is performed to determine $\phi_0(x,y)$. 

\subsection{Background generation, detection and subtraction}

\begin{figure}
    \centering
    \includegraphics[width=0.8\linewidth]{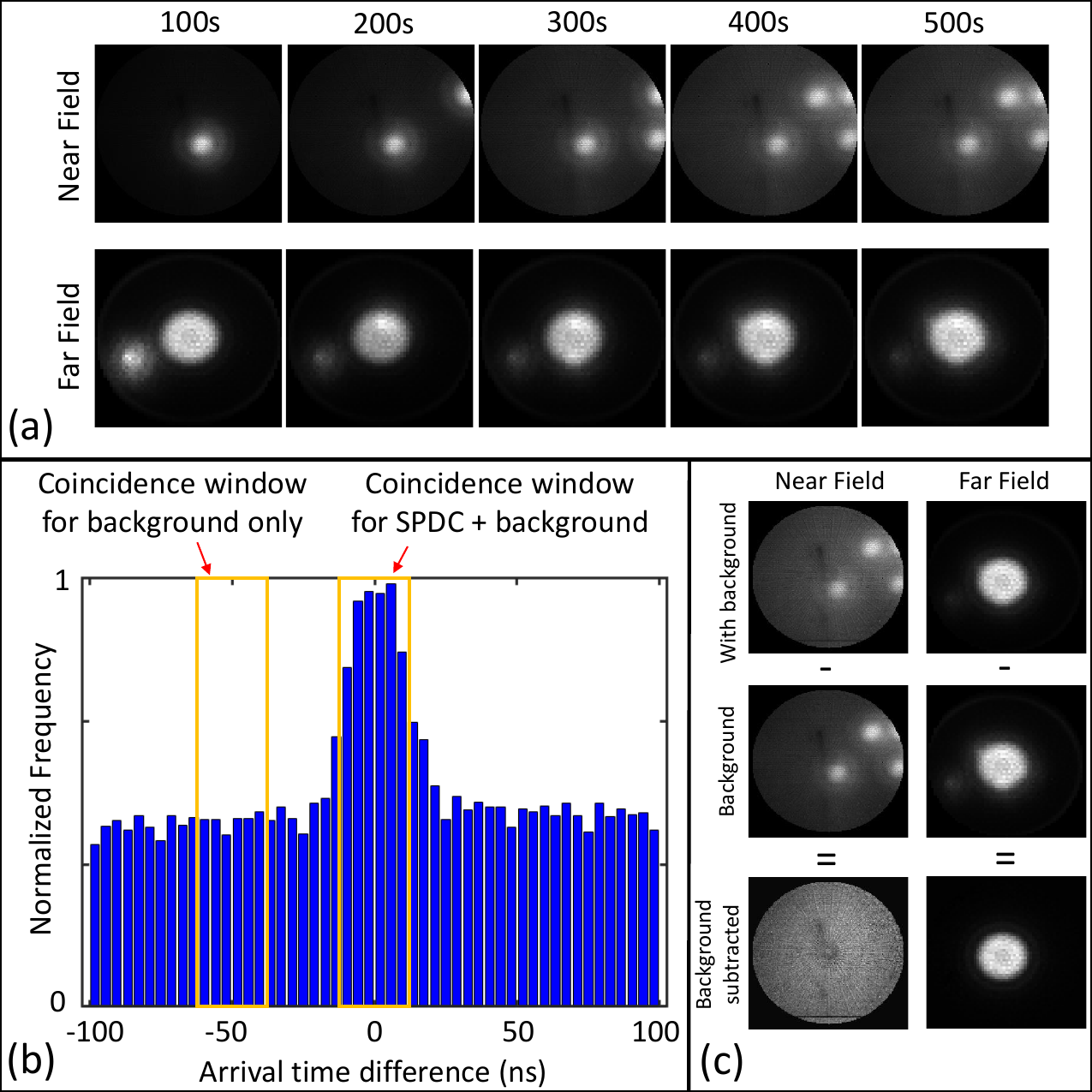}
    \caption{\textbf{Background generation and detection} (a) Shows the background laser being applied to the NF and FF of SPDC. (b) A typical coincidence histogram is shown, showing the number of events detected as a function of the difference in arrival time between the photons detected in the two regions of the camera. The location and width of the coincidence window for identifying all photons detected in coincidence and only those of background coincidences are highlighted in orange. (c) Verifying the validity of the background identification approach illustrated in (b) by subtracting the obtained background image from the image containing all coincidences.}
    \label{Method2}
\end{figure}

For creating a dynamic background light, an attenuated 780\,nm diode laser is directed into the setup at a slight angle to the SPDC as depicted in Fig.~\ref{Setup}. The number of background photons detected is approximately 67\% of SPDC. The laser beam spot was repositioned every 100\,s to simulate a dynamic background as seen in Fig.\,\ref{Method2}(a). 

Figure\,\ref{Method2}(b) is a typical histogram showing the difference in arrival time between two photons detected in the two regions of the camera. The central peak is the result of time-correlated SPDC photon pairs being detected in coincidence. On the other hand, a constant background is also visible in the histogram, which is due to the detection of coincidences between uncorrelated photons. It is not possible to determine which of the photon pairs detected are from background light; however, one can determine the resultant contribution of the background. 

To obtain the background contribution, the coincidence window is shifted away from the central peak. All photon pair events detected within this window will be from only uncorrelated photons. Since the background coincidence rate is uniform in time on the nanosecond timescale, this measurement is representative of the accidental coincidence rate in the central window. It can be verified that the resultant image formed by the uncorrelated photon pairs is indeed the background by subtracting it directly from the coincidence image to obtain a background-free image, as shown in Fig.\ref{Method2}(c). 

A 50\,ns shift in the coincidence window is used in this experiment. For a detector with better timing resolution, which will result in a narrower central peak for SPDC, a smaller shift can be used. Note that this method of background identification is only valid when the background fluctuation is slower than the applied shift to the coincidence window. 

Now, with the position information on each detected photon pair from the background correlation measurement, the background contribution is subtracted from the centroid shift as follows:
\begin{align}
\mathcal{U}(\bm{r}_s) &= \frac{\int u_i C(\bm{r}_s,\bm{k}_i) du_i - \int u_i C_B(\bm{r}_s,\bm{k}_i) du_i}{\int C(\bm{r}_s,\bm{k}_i) du_i - \int C_B(\bm{r}_s,\bm{k}_i) du_i}\nonumber\\
\mathcal{V}(\bm{r}_s) &= \frac{\int v_i C(\bm{r}_s,\bm{k}_i) dv_i - \int v_i C_B(\bm{r}_s,\bm{k}_i) dv_i}{\int C(\bm{r}_s,\bm{k}_i) dv_i - \int C_B(\bm{r}_s,\bm{k}_i) dv_i},
\end{align}
where $C_B(\bm{r}_s,\bm{k}_i)$ is the number of background coincidences detected at signal position $\bm{r}_s$ and idler momentum $\bm{k}_i$.

\newpage

\section*{Supplementary Materials}
\subsection*{The Frankot and Chellappa method}

Letting $p(x,y) = \frac{\partial\phi(x,y)}{\partial x}$ and $q(x,y) = \frac{\partial\phi(x,y)}{\partial y}$ be the measured phase gradients, the Laplacian for the phase $\phi(x,y)$ can be written as
\begin{equation}
    \nabla^2\phi(x,y) = \frac{\partial^2\phi(x,y)}{\partial x^2} + \frac{\partial^2\phi(x,y)}{\partial y^2} = \frac{\partial p(x,y)}{\partial x} + \frac{\partial q(x,y)}{\partial y}. 
\label{eq1}
\end{equation}

Given the Fourier transform and inverse Fourier transform defined as
\begin{align}
    &\mathcal{F}\Bigl[f(x,y)\Bigr] = \hat{f}(u,v) = \int\int f(x,y) e^{-i(ux+vy)} dxdy \nonumber\\
    &\mathcal{F}^{-1}\left[\hat{f}(u,v)\right] = f(x,y) = \frac{1}{2\pi}\int\int \hat{f}(u,v) e^{i(ux+vy)} dudv,
\label{eq2}
\end{align}

and using the differentiation property of the Fourier transform
\begin{equation}
    \mathcal{F}\left[\frac{d^n}{d x^n}f(x)\right] = (iu)^n\hat{f}(u),
\label{eq3}
\end{equation}

we can write the Fourier transform of Eq.\,\ref{eq1} as
\begin{align}
    \mathcal{F}\left[\frac{\partial^2\phi(x,y)}{\partial x^2} + \frac{\partial^2\phi(x,y)}{\partial y^2}\right] &= \mathcal{F}\left[\frac{\partial p(x,y)}{\partial x} + \frac{\partial q(x,y)}{\partial y}\right] \nonumber\\
    -(u^2+v^2)\hat{\phi}(u,v) &= iu\hat{p}(u,v) + iv\hat{q}(u,v).    
\label{eq4}
\end{align}

Thus, knowing the phase gradients $p(x,y)$ and $q(x,y)$ and computing their Fourier transforms $\hat{p}(u,v)$ and $\hat{q}(u,v)$, the phase $\phi(x,y)$ can be recovered through the inverse Fourier transform of Eq.\,\ref{eq4} 
\begin{equation}
    \phi(x,y) = \mathcal{F}^{-1}\left[\frac{u\hat{p}(u,v) + v\hat{q}(u,v)}{i(u^2+v^2)}\right].
\end{equation}

\subsection*{Images of full data set}
\begin{figure}[h]
    \centering
    \includegraphics[width=1\linewidth]{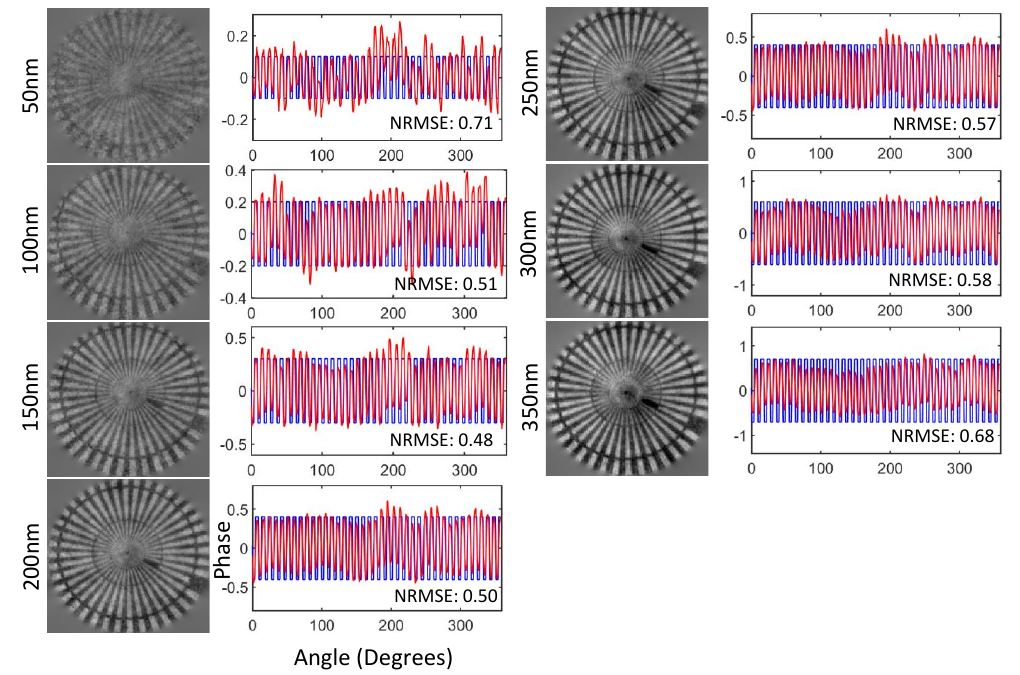}
    \caption{Recovered phase image and cross-section of Star resolution phase targets.}
    \label{Supp1}
\end{figure}

\begin{figure}[h]
    \centering
    \includegraphics[width=1\linewidth]{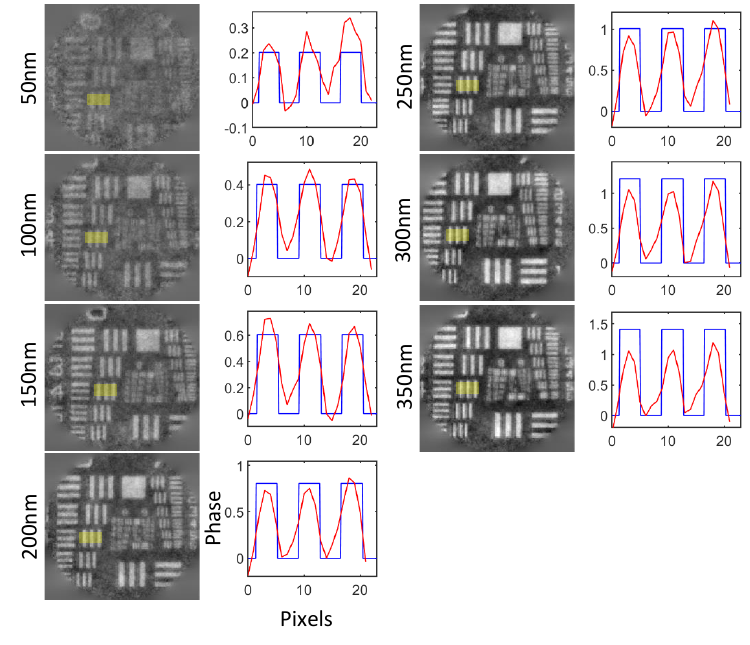}
    \caption{Recovered phase image and cross-section of the highlighted region for 1951 USAF resolution phase targets.}
    \label{Supp3}
\end{figure}

\begin{figure}[h]
    \centering
    \includegraphics[width=1\linewidth]{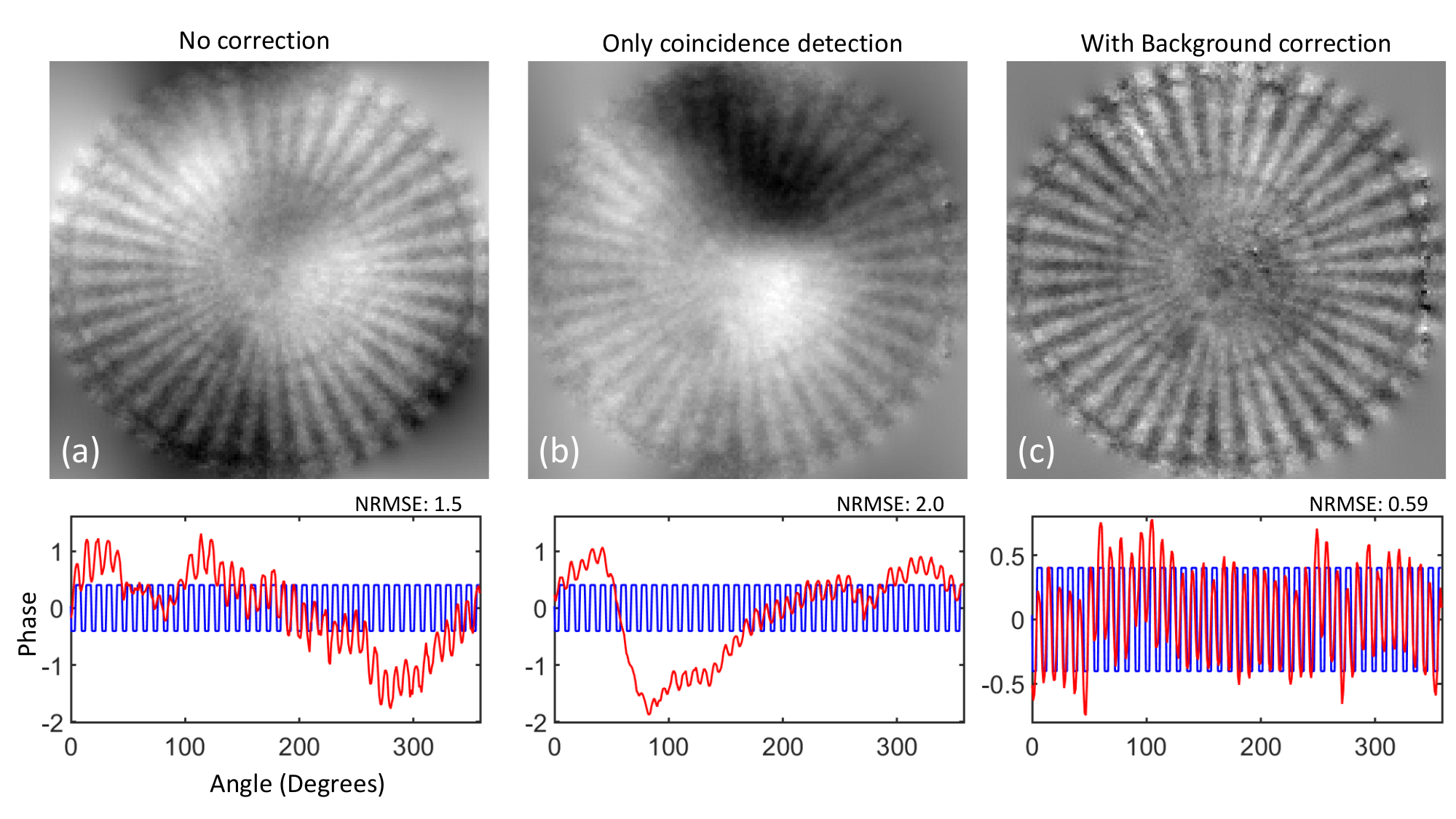}
    \caption{Recovered phase image and cross-section of Star resolution phase targets (a) without background correction, (b) only coincidence detection, (c) with coincidence and background correction.}
    \label{Supp5}
\end{figure}

Figure\,\ref{Supp1} shows the recovered phase images and cross-sections for all Star resolution phase targets. Figure\,\ref{Supp3} shows the recovered phase images and cross-sections for all 1951 USAF resolution phase targets. Figure\,\ref{Supp5} shows the recovered phase images and cross-sections of the Star resolution phase targets with and without background correction.

\subsection*{Coincidence background}
The total background in coincidence measurement $C_B$, to second order approximation is given by
\begin{equation}
    C_B = \frac{1}{T}\tau\eta_s\eta_i(P+B_s)(P+B_i),
\end{equation}
where $T$ is the data acquisition time, $\tau$ is the coincidence gating time, $\eta_{s,i}$ are the system detection efficiencies of the signal and idler photons respectively, $P$ is the number of photon pairs generated from SPDC and $B_{s,i}$ are the number of background photon in the signal and idler detectors respectively.

For simplicity we will assume $\eta_s = \eta_i = \eta$ and $B_s = B_i =B$ given that the same camera detector is used to detect both photons and the background comes mostly from the same laser.

The signal to background ratio (SBR) for coincidences is then
\begin{equation}
    \text{SBR}_C = \frac{\eta^2P}{C_B} = \frac{PT}{\tau\left(P+B\right)^2}.
\end{equation}

On the other hand, the singles SBR is simply
\begin{equation}
    \text{SBR}_S = \frac{P}{B}.
\end{equation}

In order for coincidence measurement to have a background suppression effect requires $\text{SBR}_C>\text{SBR}_S$ which requires
\begin{equation}
    P > \sqrt{BT/\tau} - B.
\end{equation}

For this experiment, when the background laser is turned on, $\tau = 20$\,ns, $T=500$\,s, $C_B \approx 7.6\times 10^6$, $\eta^2P \approx 3.9\times 10^6$ or with $\eta$ estimated to be approximately 0.015 gives $P\approx 1.7\times 10^{10}$. From these information we can determine $B\approx 1.2\times 10^{10}$ and $\sqrt{BT/\tau} - B \approx 5.4\times10^9$ which is smaller than $P$ resulting in $\text{SBR}_C<\text{SBR}_S$.

\subsection*{Comparison with a Shack-Hartmann type phase gradient microscope}
QCPGM bears many similarities to the Shack-Hartmann (SH) wavefront sensor~\cite{Shack1971}. The SH sensor employs a microlens array to capture both the position and momentum attributes of incident light rays, subsequently utilizing this data to deduce the phase information via phase gradient measurements. Though more commonly used in the field of adaptive optics~\cite{Tyson2022}, the SH sensor has also been demonstrated for use in quantitative phase microscopy~\cite{Gong2017}.  

A SH sensor requires a total of $N_n\times N_f$ camera pixels, with $N_n$ being the NF resolution given by the number of microlenses and $N_f$ the FF resolution given by the number of camera pixels behind a single microlens, which is typically around $10\times10$ pixels. The limited FF resolution gives rise to centroid pixelization error which is one of the major limiting factor in the phase measurement accuracy of a SH sensor~\cite{Neal2002,Alexander1991}. On the other hand QCPGM requires $N_n + N_f$ camera pixels which for this experiment, with a single camera of $256\times256$ pixels, we have $150\times150$ $N_n$ and $100\times100$ $N_f$ resolution. As we will see in the following simulation, with such high FF resolution, centroid pixelization error becomes negligible.  For a conventional SH sensor to have the same NF and FF resolution as the QCPGM demonstrated here, a camera of $15000\times15000$ pixels will be required, a resolution far beyond what is commercially available today for scientific cameras. Yet, higher resolution time-tagging single-photon cameras are already available which can provide an even higher resolution for QCPGM~\cite{ASI1}.

Assuming plane wave illumination, for the SH sensor, the measured FF intensity behind a microlens at position $\bm{r}_l$ can be written as
\begin{equation}
    I_\text{SH}(\bm{r}_l,\bm{k}) \propto  \Bigg|\mathcal{F} \left[T(\bm{r})\Pi\left(\frac{{\bm{r}-\bm{r}}_l}{a}\right) \right]_{\bm{r}} \Bigg|^2,
    \label{classicalI}
\end{equation}
where $\Pi(x/a)$ is the rectangular function with the width $a$ equal to the microlens diameter. Whereas the quantum scenario, the coincidence at each combination of signal photon position $\bm{r}_s$ and idler photon momentum $\bm{k}_i=(u_i,v_i)$ is given by 
\begin{equation}
    C(\bm{r}_s,\bm{k}_i) \propto  \Big|T(\bm{r}_s) \mathcal{F} \left[T(\bm{r}_i)\psi({\bm{r}}_s,{\bm{r}}_i) \right]_{\bm{r}_i} \Big|^2.
    \label{quantumI}
\end{equation}
We see that the $\Pi$ function is replaced by the biphoton correlation function. This is equivalent to having NF pixels with non-uniform, Gaussian shaped sensitivity. There is also an extra factor of $T(\bm{r}_s)$ outside the Fourier transform which gives QCPGM squared sensitivity to photon absorption.

In Fig.\ref{Fig4}, we simulate in one dimension the phase recovery accuracy of using QCPGM compared to a conventional SH sensor. A phase target is sampled with 10 NF pixels over 200\,$\mu$m for both scenarios while QCPGM has 100 pixels in the FF, and the SH sensor has 11 pixels in the FF. This is to make the fairest comparison where a total of 110 physical pixels are used for both scenarios. The target was imaged with $10^5$ photons, which are randomly distributed based on the intensity distribution given by Eqs.~\ref{classicalI} and ~\ref{quantumI}. For Eq.~\ref{quantumI}, we set the NF correlation width to be $2\delta_r = 20$\,$\mu$m, the size of a NF pixel, and assume that the second $|\bm{r}_s+\bm{r}_i|$ Gaussian term is constant within the small region we are imaging. For Eq.~\ref{classicalI}, the width of the $\Pi$ function is also set to $a = 20$\,$\mu$m. This simulation is reproduced 200 times to produce 200 images that exhibit shot noise, and that the inaccuracies on the phase due to shot noise and the resolution limits is evaluated. The resulting phase gradient and the recovered phase are shown in Fig.\ref{Fig4}(a1,a2) for the quantum scenario and Fig.\ref{Fig4}(b1,b2) for the classical SH.

The number of FF pixels is then varied for both methods to evaluate the effect of FF resolution on the measured phase accuracy and precision. The phase accuracy is evaluated with the normalized root mean squared error (NRMSE), as defined in the main text, and the uncertainty from repeated simulations evaluates the phase precision. In Fig.\ref{Fig4}(c1) we see that when compared to the ground truth, both QCPGM and SH have a sharp worsening of the NRMSE at around 20 FF pixels. This is when the FF beam width becomes comparable to that of the FF pixel size and the effect of centroid pixelization error starts setting in. Overall QCPGM has a worse NRMSE compared to a SH when both have the same FF resolution,  a direct result of the Gaussian shaped correlation function in QCPGM is averaging the phase over a wider region compared to the $\Pi$ function used for a SH. However, this difference is small at high FF resolutions and with 100 FF pixels used in QCPGM still offers a significantly better NRMSE compared to the SH at 11 FF pixels. 

In Fig.\ref{Fig4}(c2) we see that there is a small increase in the uncertainty with lower number of FF pixels for both QCPGM and SH, however, QCPGM has in overall, about 4 times lower uncertainty compared to the SH. This is a result of the wider Gaussian correlation function giving a smaller beam width in the FF compared to the $\Pi$ function thus achieving a lower centroid uncertainty, which is proportional to the beam width. 

\begin{figure}
    \centering
    \includegraphics[width=1\linewidth]{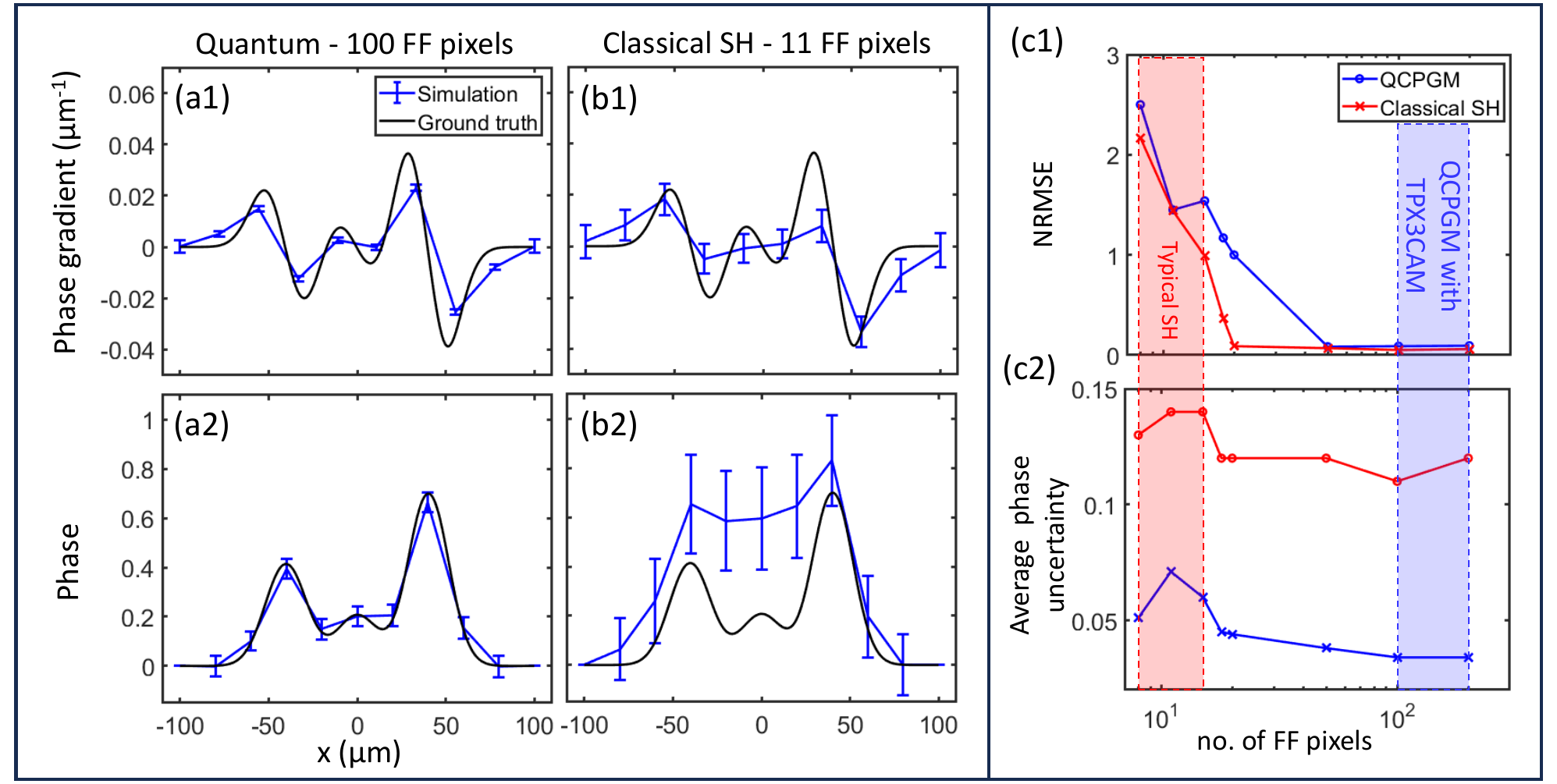}
    \caption{(a1, a2) shows the phase gradient and recovered phase using QCPGM with 10 NF pixels and 100 FF pixels. (b1, b2) is the phase gradient and recovered phase using a SH sensor with 10 NF and 11 FF pixels. Black lines are the target phase and phase gradient sampled at a resolution of 200 NF pixels which we consider to be the ground truth. Blue line is the simulated results using $10^5$ photons and is repeated 200 times to determine the uncertainties. (c1) NRMSE of the recovered phase from comparing simulation with the ground truth and (c2) the average uncertainty for QCPGM (blue line) and a SH sensor (red line) as a function of the number of FF pixels. Typical SH sensors has up to 15 FF pixels and QCPGM with the TPX3CAM can have either 100 or 200 FF pixels depending on whether one or two cameras are used.} 
    \label{Fig4}
\end{figure}

In addition to an increase in both accuracy and precision of the phase measurement, at equivalent illumination light level and number of pixel resources, QCPGM also circumvent the issue of cross-talk between pixels situated behind neighboring microlenses, a common problem encountered in SH sensors when encountering large phase gradient which could lead to wildly fluctuating phase gradients. Lastly, QCPGM employs standard 1- and 2-inch diameter lenses, which can be constructed with much higher quality compared to the less-ideal shape of microlens arrays and is less prone to misalignment, thus reducing the occurrence of aberrations~\cite{Pfund1998,Chernyshov2005}. 

\bibliography{QQPIref}

\begin{thebibliography}{67}%
\makeatletter
\providecommand \@ifxundefined [1]{%
 \@ifx{#1\undefined}
}%
\providecommand \@ifnum [1]{%
 \ifnum #1\expandafter \@firstoftwo
 \else \expandafter \@secondoftwo
 \fi
}%
\providecommand \@ifx [1]{%
 \ifx #1\expandafter \@firstoftwo
 \else \expandafter \@secondoftwo
 \fi
}%
\providecommand \natexlab [1]{#1}%
\providecommand \enquote  [1]{``#1''}%
\providecommand \bibnamefont  [1]{#1}%
\providecommand \bibfnamefont [1]{#1}%
\providecommand \citenamefont [1]{#1}%
\providecommand \href@noop [0]{\@secondoftwo}%
\providecommand \href [0]{\begingroup \@sanitize@url \@href}%
\providecommand \@href[1]{\@@startlink{#1}\@@href}%
\providecommand \@@href[1]{\endgroup#1\@@endlink}%
\providecommand \@sanitize@url [0]{\catcode `\\12\catcode `\$12\catcode
  `\&12\catcode `\#12\catcode `\^12\catcode `\_12\catcode `\%12\relax}%
\providecommand \@@startlink[1]{}%
\providecommand \@@endlink[0]{}%
\providecommand \url  [0]{\begingroup\@sanitize@url \@url }%
\providecommand \@url [1]{\endgroup\@href {#1}{\urlprefix }}%
\providecommand \urlprefix  [0]{URL }%
\providecommand \Eprint [0]{\href }%
\providecommand \doibase [0]{https://doi.org/}%
\providecommand \selectlanguage [0]{\@gobble}%
\providecommand \bibinfo  [0]{\@secondoftwo}%
\providecommand \bibfield  [0]{\@secondoftwo}%
\providecommand \translation [1]{[#1]}%
\providecommand \BibitemOpen [0]{}%
\providecommand \bibitemStop [0]{}%
\providecommand \bibitemNoStop [0]{.\EOS\space}%
\providecommand \EOS [0]{\spacefactor3000\relax}%
\providecommand \BibitemShut  [1]{\csname bibitem#1\endcsname}%
\let\auto@bib@innerbib\@empty
\bibitem [{\citenamefont {Zernike}(1942)}]{Zernike1}%
  \BibitemOpen
  \bibfield  {author} {\bibinfo {author} {\bibfnamefont {F.}~\bibnamefont
  {Zernike}},\ }\bibfield  {title} {\bibinfo {title} {Phase contrast, a new
  method for the microscopic observation of transparent objects},\ }\href
  {https://doi.org/https://doi.org/10.1016/S0031-8914(42)80035-X} {\bibfield
  {journal} {\bibinfo  {journal} {Physica}\ }\textbf {\bibinfo {volume} {9}},\
  \bibinfo {pages} {686} (\bibinfo {year} {1942})}\BibitemShut {NoStop}%
\bibitem [{\citenamefont {Zernike}(1955)}]{Zernike2}%
  \BibitemOpen
  \bibfield  {author} {\bibinfo {author} {\bibfnamefont {F.}~\bibnamefont
  {Zernike}},\ }\bibfield  {title} {\bibinfo {title} {How i discovered phase
  contrast},\ }\href {https://doi.org/10.1126/science.121.3141.345} {\bibfield
  {journal} {\bibinfo  {journal} {Science}\ }\textbf {\bibinfo {volume}
  {121}},\ \bibinfo {pages} {345} (\bibinfo {year} {1955})},\ \Eprint
  {https://arxiv.org/abs/https://www.science.org/doi/pdf/10.1126/science.121.3141.345}
  {https://www.science.org/doi/pdf/10.1126/science.121.3141.345} \BibitemShut
  {NoStop}%
\bibitem [{\citenamefont {Nguyen}\ \emph {et~al.}(2022)\citenamefont {Nguyen},
  \citenamefont {Pradeep}, \citenamefont {Judson-Torres}, \citenamefont {Reed},
  \citenamefont {Teitell},\ and\ \citenamefont {Zangle}}]{Nguyen2022}%
  \BibitemOpen
  \bibfield  {author} {\bibinfo {author} {\bibfnamefont {T.~L.}\ \bibnamefont
  {Nguyen}}, \bibinfo {author} {\bibfnamefont {S.}~\bibnamefont {Pradeep}},
  \bibinfo {author} {\bibfnamefont {R.~L.}\ \bibnamefont {Judson-Torres}},
  \bibinfo {author} {\bibfnamefont {J.}~\bibnamefont {Reed}}, \bibinfo {author}
  {\bibfnamefont {M.~A.}\ \bibnamefont {Teitell}},\ and\ \bibinfo {author}
  {\bibfnamefont {T.~A.}\ \bibnamefont {Zangle}},\ }\bibfield  {title}
  {\bibinfo {title} {Quantitative phase imaging: Recent advances and expanding
  potential in biomedicine},\ }\href {https://doi.org/10.1021/acsnano.1c11507}
  {\bibfield  {journal} {\bibinfo  {journal} {ACS Nano}\ }\textbf {\bibinfo
  {volume} {16}},\ \bibinfo {pages} {11516} (\bibinfo {year} {2022})},\
  \bibinfo {note} {pMID: 35916417}\BibitemShut {NoStop}%
\bibitem [{\citenamefont {Chaumet}\ \emph {et~al.}(2024)\citenamefont
  {Chaumet}, \citenamefont {Bon}, \citenamefont {Maire}, \citenamefont
  {Sentenac},\ and\ \citenamefont {Baffou}}]{Chaumet2024}%
  \BibitemOpen
  \bibfield  {author} {\bibinfo {author} {\bibfnamefont {P.~C.}\ \bibnamefont
  {Chaumet}}, \bibinfo {author} {\bibfnamefont {P.}~\bibnamefont {Bon}},
  \bibinfo {author} {\bibfnamefont {G.}~\bibnamefont {Maire}}, \bibinfo
  {author} {\bibfnamefont {A.}~\bibnamefont {Sentenac}},\ and\ \bibinfo
  {author} {\bibfnamefont {G.}~\bibnamefont {Baffou}},\ }\bibfield  {title}
  {\bibinfo {title} {Quantitative phase microscopies: accuracy comparison},\
  }\href {https://doi.org/10.1038/s41377-024-01619-7} {\bibfield  {journal}
  {\bibinfo  {journal} {Light: Science {\&} Applications}\ }\textbf {\bibinfo
  {volume} {13}},\ \bibinfo {pages} {288} (\bibinfo {year} {2024})}\BibitemShut
  {NoStop}%
\bibitem [{\citenamefont {Huang}\ and\ \citenamefont {Cao}(2024)}]{Huang2024}%
  \BibitemOpen
  \bibfield  {author} {\bibinfo {author} {\bibfnamefont {Z.}~\bibnamefont
  {Huang}}\ and\ \bibinfo {author} {\bibfnamefont {L.}~\bibnamefont {Cao}},\
  }\bibfield  {title} {\bibinfo {title} {Quantitative phase imaging based on
  holography: trends and new perspectives},\ }\href
  {https://doi.org/10.1038/s41377-024-01453-x} {\bibfield  {journal} {\bibinfo
  {journal} {Light: Science {\&} Applications}\ }\textbf {\bibinfo {volume}
  {13}},\ \bibinfo {pages} {145} (\bibinfo {year} {2024})}\BibitemShut
  {NoStop}%
\bibitem [{\citenamefont {Shack}(1971)}]{Shack1971}%
  \BibitemOpen
  \bibfield  {author} {\bibinfo {author} {\bibfnamefont {R.~V.}\ \bibnamefont
  {Shack}},\ }\bibfield  {title} {\bibinfo {title} {Production and use of a
  lenticular hartmann screen},\ }\href
  {https://cir.nii.ac.jp/crid/1573387450670986240} {\bibfield  {journal}
  {\bibinfo  {journal} {Spring Meeting of Optical Society of America, 1971}\
  }\textbf {\bibinfo {volume} {656}} (\bibinfo {year} {1971})}\BibitemShut
  {NoStop}%
\bibitem [{\citenamefont {Gong}\ \emph {et~al.}(2017)\citenamefont {Gong},
  \citenamefont {Agbana}, \citenamefont {Pozzi}, \citenamefont {Soloviev},
  \citenamefont {Verhaegen},\ and\ \citenamefont {Vdovin}}]{Gong2017}%
  \BibitemOpen
  \bibfield  {author} {\bibinfo {author} {\bibfnamefont {H.}~\bibnamefont
  {Gong}}, \bibinfo {author} {\bibfnamefont {T.~E.}\ \bibnamefont {Agbana}},
  \bibinfo {author} {\bibfnamefont {P.}~\bibnamefont {Pozzi}}, \bibinfo
  {author} {\bibfnamefont {O.}~\bibnamefont {Soloviev}}, \bibinfo {author}
  {\bibfnamefont {M.}~\bibnamefont {Verhaegen}},\ and\ \bibinfo {author}
  {\bibfnamefont {G.}~\bibnamefont {Vdovin}},\ }\bibfield  {title} {\bibinfo
  {title} {Optical path difference microscopy with a shack--hartmann wavefront
  sensor},\ }\href {https://doi.org/10.1364/OL.42.002122} {\bibfield  {journal}
  {\bibinfo  {journal} {Opt. Lett.}\ }\textbf {\bibinfo {volume} {42}},\
  \bibinfo {pages} {2122} (\bibinfo {year} {2017})}\BibitemShut {NoStop}%
\bibitem [{\citenamefont {Streibl}(1984)}]{Streibl1984}%
  \BibitemOpen
  \bibfield  {author} {\bibinfo {author} {\bibfnamefont {N.}~\bibnamefont
  {Streibl}},\ }\bibfield  {title} {\bibinfo {title} {Phase imaging by the
  transport equation of intensity},\ }\href
  {https://doi.org/https://doi.org/10.1016/0030-4018(84)90079-8} {\bibfield
  {journal} {\bibinfo  {journal} {Optics Communications}\ }\textbf {\bibinfo
  {volume} {49}},\ \bibinfo {pages} {6} (\bibinfo {year} {1984})}\BibitemShut
  {NoStop}%
\bibitem [{\citenamefont {Gerchberg}(1972)}]{Gerchberg1972}%
  \BibitemOpen
  \bibfield  {author} {\bibinfo {author} {\bibfnamefont {R.~W.}\ \bibnamefont
  {Gerchberg}},\ }\bibfield  {title} {\bibinfo {title} {A practical algorithm
  for the determination of phase from image and diffraction plane pictures},\
  }\href@noop {} {\bibfield  {journal} {\bibinfo  {journal} {Optik}\ }\textbf
  {\bibinfo {volume} {35}},\ \bibinfo {pages} {237} (\bibinfo {year}
  {1972})}\BibitemShut {NoStop}%
\bibitem [{\citenamefont {Fienup}(1982)}]{Fienup1982}%
  \BibitemOpen
  \bibfield  {author} {\bibinfo {author} {\bibfnamefont {J.~R.}\ \bibnamefont
  {Fienup}},\ }\bibfield  {title} {\bibinfo {title} {Phase retrieval
  algorithms: a comparison},\ }\href {https://doi.org/10.1364/AO.21.002758}
  {\bibfield  {journal} {\bibinfo  {journal} {Appl. Opt.}\ }\textbf {\bibinfo
  {volume} {21}},\ \bibinfo {pages} {2758} (\bibinfo {year}
  {1982})}\BibitemShut {NoStop}%
\bibitem [{\citenamefont {Zheng}\ \emph {et~al.}(2013)\citenamefont {Zheng},
  \citenamefont {Horstmeyer},\ and\ \citenamefont {Yang}}]{Zheng2013}%
  \BibitemOpen
  \bibfield  {author} {\bibinfo {author} {\bibfnamefont {G.}~\bibnamefont
  {Zheng}}, \bibinfo {author} {\bibfnamefont {R.}~\bibnamefont {Horstmeyer}},\
  and\ \bibinfo {author} {\bibfnamefont {C.}~\bibnamefont {Yang}},\ }\bibfield
  {title} {\bibinfo {title} {Wide-field, high-resolution fourier ptychographic
  microscopy},\ }\href {https://doi.org/10.1038/nphoton.2013.187} {\bibfield
  {journal} {\bibinfo  {journal} {Nature Photonics}\ }\textbf {\bibinfo
  {volume} {7}},\ \bibinfo {pages} {739} (\bibinfo {year} {2013})}\BibitemShut
  {NoStop}%
\bibitem [{\citenamefont {Tian}\ and\ \citenamefont
  {Waller}(2015{\natexlab{a}})}]{Tian2015}%
  \BibitemOpen
  \bibfield  {author} {\bibinfo {author} {\bibfnamefont {L.}~\bibnamefont
  {Tian}}\ and\ \bibinfo {author} {\bibfnamefont {L.}~\bibnamefont {Waller}},\
  }\bibfield  {title} {\bibinfo {title} {3d intensity and phase imaging from
  light field measurements in an led array microscope},\ }\href
  {https://doi.org/10.1364/OPTICA.2.000104} {\bibfield  {journal} {\bibinfo
  {journal} {Optica}\ }\textbf {\bibinfo {volume} {2}},\ \bibinfo {pages} {104}
  (\bibinfo {year} {2015}{\natexlab{a}})}\BibitemShut {NoStop}%
\bibitem [{\citenamefont {Tian}\ and\ \citenamefont
  {Waller}(2015{\natexlab{b}})}]{Tian2015_2}%
  \BibitemOpen
  \bibfield  {author} {\bibinfo {author} {\bibfnamefont {L.}~\bibnamefont
  {Tian}}\ and\ \bibinfo {author} {\bibfnamefont {L.}~\bibnamefont {Waller}},\
  }\bibfield  {title} {\bibinfo {title} {Quantitative differential phase
  contrast imaging in an led array microscope},\ }\href
  {https://doi.org/10.1364/OE.23.011394} {\bibfield  {journal} {\bibinfo
  {journal} {Opt. Express}\ }\textbf {\bibinfo {volume} {23}},\ \bibinfo
  {pages} {11394} (\bibinfo {year} {2015}{\natexlab{b}})}\BibitemShut {NoStop}%
\bibitem [{\citenamefont {Mehta}\ and\ \citenamefont
  {Sheppard}(2009)}]{Mehta2009}%
  \BibitemOpen
  \bibfield  {author} {\bibinfo {author} {\bibfnamefont {S.~B.}\ \bibnamefont
  {Mehta}}\ and\ \bibinfo {author} {\bibfnamefont {C.~J.~R.}\ \bibnamefont
  {Sheppard}},\ }\bibfield  {title} {\bibinfo {title} {Quantitative
  phase-gradient imaging at high resolution with asymmetric illumination-based
  differential phase contrast},\ }\href {https://doi.org/10.1364/OL.34.001924}
  {\bibfield  {journal} {\bibinfo  {journal} {Opt. Lett.}\ }\textbf {\bibinfo
  {volume} {34}},\ \bibinfo {pages} {1924} (\bibinfo {year}
  {2009})}\BibitemShut {NoStop}%
\bibitem [{\citenamefont {Genovese}(2016)}]{Genovese_2016}%
  \BibitemOpen
  \bibfield  {author} {\bibinfo {author} {\bibfnamefont {M.}~\bibnamefont
  {Genovese}},\ }\bibfield  {title} {\bibinfo {title} {Real applications of
  quantum imaging},\ }\href {https://doi.org/10.1088/2040-8978/18/7/073002}
  {\bibfield  {journal} {\bibinfo  {journal} {Journal of Optics}\ }\textbf
  {\bibinfo {volume} {18}},\ \bibinfo {pages} {073002} (\bibinfo {year}
  {2016})}\BibitemShut {NoStop}%
\bibitem [{\citenamefont {Berchera}\ and\ \citenamefont
  {Degiovanni}(2019)}]{Berchera_2019}%
  \BibitemOpen
  \bibfield  {author} {\bibinfo {author} {\bibfnamefont {I.~R.}\ \bibnamefont
  {Berchera}}\ and\ \bibinfo {author} {\bibfnamefont {I.~P.}\ \bibnamefont
  {Degiovanni}},\ }\bibfield  {title} {\bibinfo {title} {Quantum imaging with
  sub-poissonian light: challenges and perspectives in optical metrology},\
  }\href {https://doi.org/10.1088/1681-7575/aaf7b2} {\bibfield  {journal}
  {\bibinfo  {journal} {Metrologia}\ }\textbf {\bibinfo {volume} {56}},\
  \bibinfo {pages} {024001} (\bibinfo {year} {2019})}\BibitemShut {NoStop}%
\bibitem [{\citenamefont {Moreau}\ \emph {et~al.}(2019)\citenamefont {Moreau},
  \citenamefont {Toninelli}, \citenamefont {Gregory},\ and\ \citenamefont
  {Padgett}}]{Moreau2019}%
  \BibitemOpen
  \bibfield  {author} {\bibinfo {author} {\bibfnamefont {P.-A.}\ \bibnamefont
  {Moreau}}, \bibinfo {author} {\bibfnamefont {E.}~\bibnamefont {Toninelli}},
  \bibinfo {author} {\bibfnamefont {T.}~\bibnamefont {Gregory}},\ and\ \bibinfo
  {author} {\bibfnamefont {M.~J.}\ \bibnamefont {Padgett}},\ }\bibfield
  {title} {\bibinfo {title} {Imaging with quantum states of light},\ }\href
  {https://doi.org/10.1038/s42254-019-0056-0} {\bibfield  {journal} {\bibinfo
  {journal} {Nature Reviews Physics}\ }\textbf {\bibinfo {volume} {1}},\
  \bibinfo {pages} {367} (\bibinfo {year} {2019})}\BibitemShut {NoStop}%
\bibitem [{\citenamefont {Degen}\ \emph {et~al.}(2017)\citenamefont {Degen},
  \citenamefont {Reinhard},\ and\ \citenamefont {Cappellaro}}]{Degen2017}%
  \BibitemOpen
  \bibfield  {author} {\bibinfo {author} {\bibfnamefont {C.~L.}\ \bibnamefont
  {Degen}}, \bibinfo {author} {\bibfnamefont {F.}~\bibnamefont {Reinhard}},\
  and\ \bibinfo {author} {\bibfnamefont {P.}~\bibnamefont {Cappellaro}},\
  }\bibfield  {title} {\bibinfo {title} {Quantum sensing},\ }\href
  {https://doi.org/10.1103/RevModPhys.89.035002} {\bibfield  {journal}
  {\bibinfo  {journal} {Rev. Mod. Phys.}\ }\textbf {\bibinfo {volume} {89}},\
  \bibinfo {pages} {035002} (\bibinfo {year} {2017})}\BibitemShut {NoStop}%
\bibitem [{\citenamefont {Pirandola}\ \emph {et~al.}(2018)\citenamefont
  {Pirandola}, \citenamefont {Bardhan}, \citenamefont {Gehring}, \citenamefont
  {Weedbrook},\ and\ \citenamefont {Lloyd}}]{Pirandola2018}%
  \BibitemOpen
  \bibfield  {author} {\bibinfo {author} {\bibfnamefont {S.}~\bibnamefont
  {Pirandola}}, \bibinfo {author} {\bibfnamefont {B.~R.}\ \bibnamefont
  {Bardhan}}, \bibinfo {author} {\bibfnamefont {T.}~\bibnamefont {Gehring}},
  \bibinfo {author} {\bibfnamefont {C.}~\bibnamefont {Weedbrook}},\ and\
  \bibinfo {author} {\bibfnamefont {S.}~\bibnamefont {Lloyd}},\ }\bibfield
  {title} {\bibinfo {title} {Advances in photonic quantum sensing},\ }\href
  {https://doi.org/10.1038/s41566-018-0301-6} {\bibfield  {journal} {\bibinfo
  {journal} {Nature Photonics}\ }\textbf {\bibinfo {volume} {12}},\ \bibinfo
  {pages} {724} (\bibinfo {year} {2018})}\BibitemShut {NoStop}%
\bibitem [{\citenamefont {Tsang}(2009)}]{Tsang2009}%
  \BibitemOpen
  \bibfield  {author} {\bibinfo {author} {\bibfnamefont {M.}~\bibnamefont
  {Tsang}},\ }\bibfield  {title} {\bibinfo {title} {Quantum imaging beyond the
  diffraction limit by optical centroid measurements},\ }\href
  {https://doi.org/10.1103/PhysRevLett.102.253601} {\bibfield  {journal}
  {\bibinfo  {journal} {Phys. Rev. Lett.}\ }\textbf {\bibinfo {volume} {102}},\
  \bibinfo {pages} {253601} (\bibinfo {year} {2009})}\BibitemShut {NoStop}%
\bibitem [{\citenamefont {Shin}\ \emph {et~al.}(2011)\citenamefont {Shin},
  \citenamefont {Chan}, \citenamefont {Chang},\ and\ \citenamefont
  {Boyd}}]{Shin2011}%
  \BibitemOpen
  \bibfield  {author} {\bibinfo {author} {\bibfnamefont {H.}~\bibnamefont
  {Shin}}, \bibinfo {author} {\bibfnamefont {K.~W.~C.}\ \bibnamefont {Chan}},
  \bibinfo {author} {\bibfnamefont {H.~J.}\ \bibnamefont {Chang}},\ and\
  \bibinfo {author} {\bibfnamefont {R.~W.}\ \bibnamefont {Boyd}},\ }\bibfield
  {title} {\bibinfo {title} {Quantum spatial superresolution by optical
  centroid measurements},\ }\href
  {https://doi.org/10.1103/PhysRevLett.107.083603} {\bibfield  {journal}
  {\bibinfo  {journal} {Phys. Rev. Lett.}\ }\textbf {\bibinfo {volume} {107}},\
  \bibinfo {pages} {083603} (\bibinfo {year} {2011})}\BibitemShut {NoStop}%
\bibitem [{\citenamefont {Rozema}\ \emph {et~al.}(2014)\citenamefont {Rozema},
  \citenamefont {Bateman}, \citenamefont {Mahler}, \citenamefont {Okamoto},
  \citenamefont {Feizpour}, \citenamefont {Hayat},\ and\ \citenamefont
  {Steinberg}}]{Rozema2014}%
  \BibitemOpen
  \bibfield  {author} {\bibinfo {author} {\bibfnamefont {L.~A.}\ \bibnamefont
  {Rozema}}, \bibinfo {author} {\bibfnamefont {J.~D.}\ \bibnamefont {Bateman}},
  \bibinfo {author} {\bibfnamefont {D.~H.}\ \bibnamefont {Mahler}}, \bibinfo
  {author} {\bibfnamefont {R.}~\bibnamefont {Okamoto}}, \bibinfo {author}
  {\bibfnamefont {A.}~\bibnamefont {Feizpour}}, \bibinfo {author}
  {\bibfnamefont {A.}~\bibnamefont {Hayat}},\ and\ \bibinfo {author}
  {\bibfnamefont {A.~M.}\ \bibnamefont {Steinberg}},\ }\bibfield  {title}
  {\bibinfo {title} {Scalable spatial superresolution using entangled
  photons},\ }\href {https://doi.org/10.1103/PhysRevLett.112.223602} {\bibfield
   {journal} {\bibinfo  {journal} {Phys. Rev. Lett.}\ }\textbf {\bibinfo
  {volume} {112}},\ \bibinfo {pages} {223602} (\bibinfo {year}
  {2014})}\BibitemShut {NoStop}%
\bibitem [{\citenamefont {Untern\"{a}hrer}\ \emph {et~al.}(2018)\citenamefont
  {Untern\"{a}hrer}, \citenamefont {Bessire}, \citenamefont {Gasparini},
  \citenamefont {Perenzoni},\ and\ \citenamefont {Stefanov}}]{Unternahrer2018}%
  \BibitemOpen
  \bibfield  {author} {\bibinfo {author} {\bibfnamefont {M.}~\bibnamefont
  {Untern\"{a}hrer}}, \bibinfo {author} {\bibfnamefont {B.}~\bibnamefont
  {Bessire}}, \bibinfo {author} {\bibfnamefont {L.}~\bibnamefont {Gasparini}},
  \bibinfo {author} {\bibfnamefont {M.}~\bibnamefont {Perenzoni}},\ and\
  \bibinfo {author} {\bibfnamefont {A.}~\bibnamefont {Stefanov}},\ }\bibfield
  {title} {\bibinfo {title} {Super-resolution quantum imaging at the heisenberg
  limit},\ }\href {https://doi.org/10.1364/OPTICA.5.001150} {\bibfield
  {journal} {\bibinfo  {journal} {Optica}\ }\textbf {\bibinfo {volume} {5}},\
  \bibinfo {pages} {1150} (\bibinfo {year} {2018})}\BibitemShut {NoStop}%
\bibitem [{\citenamefont {Tenne}\ \emph {et~al.}(2019)\citenamefont {Tenne},
  \citenamefont {Rossman}, \citenamefont {Rephael}, \citenamefont {Israel},
  \citenamefont {Krupinski-Ptaszek}, \citenamefont {Lapkiewicz}, \citenamefont
  {Silberberg},\ and\ \citenamefont {Oron}}]{Tenne2019}%
  \BibitemOpen
  \bibfield  {author} {\bibinfo {author} {\bibfnamefont {R.}~\bibnamefont
  {Tenne}}, \bibinfo {author} {\bibfnamefont {U.}~\bibnamefont {Rossman}},
  \bibinfo {author} {\bibfnamefont {B.}~\bibnamefont {Rephael}}, \bibinfo
  {author} {\bibfnamefont {Y.}~\bibnamefont {Israel}}, \bibinfo {author}
  {\bibfnamefont {A.}~\bibnamefont {Krupinski-Ptaszek}}, \bibinfo {author}
  {\bibfnamefont {R.}~\bibnamefont {Lapkiewicz}}, \bibinfo {author}
  {\bibfnamefont {Y.}~\bibnamefont {Silberberg}},\ and\ \bibinfo {author}
  {\bibfnamefont {D.}~\bibnamefont {Oron}},\ }\bibfield  {title} {\bibinfo
  {title} {Super-resolution enhancement by quantum image scanning microscopy},\
  }\href {https://doi.org/10.1038/s41566-018-0324-z} {\bibfield  {journal}
  {\bibinfo  {journal} {Nature Photonics}\ }\textbf {\bibinfo {volume} {13}},\
  \bibinfo {pages} {116} (\bibinfo {year} {2019})}\BibitemShut {NoStop}%
\bibitem [{\citenamefont {Brida}\ \emph {et~al.}(2010)\citenamefont {Brida},
  \citenamefont {Genovese},\ and\ \citenamefont {Ruo~Berchera}}]{Brida2010}%
  \BibitemOpen
  \bibfield  {author} {\bibinfo {author} {\bibfnamefont {G.}~\bibnamefont
  {Brida}}, \bibinfo {author} {\bibfnamefont {M.}~\bibnamefont {Genovese}},\
  and\ \bibinfo {author} {\bibfnamefont {I.}~\bibnamefont {Ruo~Berchera}},\
  }\bibfield  {title} {\bibinfo {title} {Experimental realization of
  sub-shot-noise quantum imaging},\ }\href
  {https://doi.org/10.1038/nphoton.2010.29} {\bibfield  {journal} {\bibinfo
  {journal} {Nature Photonics}\ }\textbf {\bibinfo {volume} {4}},\ \bibinfo
  {pages} {227} (\bibinfo {year} {2010})}\BibitemShut {NoStop}%
\bibitem [{\citenamefont {Samantaray}\ \emph {et~al.}(2017)\citenamefont
  {Samantaray}, \citenamefont {Ruo-Berchera}, \citenamefont {Meda},\ and\
  \citenamefont {Genovese}}]{Samantaray2017}%
  \BibitemOpen
  \bibfield  {author} {\bibinfo {author} {\bibfnamefont {N.}~\bibnamefont
  {Samantaray}}, \bibinfo {author} {\bibfnamefont {I.}~\bibnamefont
  {Ruo-Berchera}}, \bibinfo {author} {\bibfnamefont {A.}~\bibnamefont {Meda}},\
  and\ \bibinfo {author} {\bibfnamefont {M.}~\bibnamefont {Genovese}},\
  }\bibfield  {title} {\bibinfo {title} {Realization of the first
  sub-shot-noise wide field micscope},\ }\href
  {https://doi.org/10.1038/lsa.2017.5} {\bibfield  {journal} {\bibinfo
  {journal} {Light: Science \& Applications}\ }\textbf {\bibinfo {volume}
  {6}},\ \bibinfo {pages} {e17005} (\bibinfo {year} {2017})}\BibitemShut
  {NoStop}%
\bibitem [{\citenamefont {Zhang}\ \emph {et~al.}(2020)\citenamefont {Zhang},
  \citenamefont {England}, \citenamefont {Nomerotski}, \citenamefont {Svihra},
  \citenamefont {Ferrante}, \citenamefont {Hockett},\ and\ \citenamefont
  {Sussman}}]{Zhang2020}%
  \BibitemOpen
  \bibfield  {author} {\bibinfo {author} {\bibfnamefont {Y.}~\bibnamefont
  {Zhang}}, \bibinfo {author} {\bibfnamefont {D.}~\bibnamefont {England}},
  \bibinfo {author} {\bibfnamefont {A.}~\bibnamefont {Nomerotski}}, \bibinfo
  {author} {\bibfnamefont {P.}~\bibnamefont {Svihra}}, \bibinfo {author}
  {\bibfnamefont {S.}~\bibnamefont {Ferrante}}, \bibinfo {author}
  {\bibfnamefont {P.}~\bibnamefont {Hockett}},\ and\ \bibinfo {author}
  {\bibfnamefont {B.}~\bibnamefont {Sussman}},\ }\bibfield  {title} {\bibinfo
  {title} {Multidimensional quantum-enhanced target detection via
  spectrotemporal-correlation measurements},\ }\href
  {https://doi.org/10.1103/PhysRevA.101.053808} {\bibfield  {journal} {\bibinfo
   {journal} {Phys. Rev. A}\ }\textbf {\bibinfo {volume} {101}},\ \bibinfo
  {pages} {053808} (\bibinfo {year} {2020})}\BibitemShut {NoStop}%
\bibitem [{\citenamefont {Defienne}\ \emph
  {et~al.}(2021{\natexlab{a}})\citenamefont {Defienne}, \citenamefont {Zhao},
  \citenamefont {Charbon},\ and\ \citenamefont {Faccio}}]{Defienne2021_2}%
  \BibitemOpen
  \bibfield  {author} {\bibinfo {author} {\bibfnamefont {H.}~\bibnamefont
  {Defienne}}, \bibinfo {author} {\bibfnamefont {J.}~\bibnamefont {Zhao}},
  \bibinfo {author} {\bibfnamefont {E.}~\bibnamefont {Charbon}},\ and\ \bibinfo
  {author} {\bibfnamefont {D.}~\bibnamefont {Faccio}},\ }\bibfield  {title}
  {\bibinfo {title} {Full-field quantum imaging with a single-photon avalanche
  diode camera},\ }\href {https://doi.org/10.1103/PhysRevA.103.042608}
  {\bibfield  {journal} {\bibinfo  {journal} {Phys. Rev. A}\ }\textbf {\bibinfo
  {volume} {103}},\ \bibinfo {pages} {042608} (\bibinfo {year}
  {2021}{\natexlab{a}})}\BibitemShut {NoStop}%
\bibitem [{\citenamefont {Zhao}\ \emph {et~al.}(2022)\citenamefont {Zhao},
  \citenamefont {Lyons}, \citenamefont {Ulku}, \citenamefont {Defienne},
  \citenamefont {Faccio},\ and\ \citenamefont {Charbon}}]{Zhao2022}%
  \BibitemOpen
  \bibfield  {author} {\bibinfo {author} {\bibfnamefont {J.}~\bibnamefont
  {Zhao}}, \bibinfo {author} {\bibfnamefont {A.}~\bibnamefont {Lyons}},
  \bibinfo {author} {\bibfnamefont {A.~C.}\ \bibnamefont {Ulku}}, \bibinfo
  {author} {\bibfnamefont {H.}~\bibnamefont {Defienne}}, \bibinfo {author}
  {\bibfnamefont {D.}~\bibnamefont {Faccio}},\ and\ \bibinfo {author}
  {\bibfnamefont {E.}~\bibnamefont {Charbon}},\ }\bibfield  {title} {\bibinfo
  {title} {Light detection and ranging with entangled photons},\ }\href
  {https://doi.org/10.1364/OE.435898} {\bibfield  {journal} {\bibinfo
  {journal} {Opt. Express}\ }\textbf {\bibinfo {volume} {30}},\ \bibinfo
  {pages} {3675} (\bibinfo {year} {2022})}\BibitemShut {NoStop}%
\bibitem [{\citenamefont {Aspden}\ \emph {et~al.}(2015)\citenamefont {Aspden},
  \citenamefont {Gemmell}, \citenamefont {Morris}, \citenamefont {Tasca},
  \citenamefont {Mertens}, \citenamefont {Tanner}, \citenamefont {Kirkwood},
  \citenamefont {Ruggeri}, \citenamefont {Tosi}, \citenamefont {Boyd},
  \citenamefont {Buller}, \citenamefont {Hadfield},\ and\ \citenamefont
  {Padgett}}]{Aspden2015}%
  \BibitemOpen
  \bibfield  {author} {\bibinfo {author} {\bibfnamefont {R.~S.}\ \bibnamefont
  {Aspden}}, \bibinfo {author} {\bibfnamefont {N.~R.}\ \bibnamefont {Gemmell}},
  \bibinfo {author} {\bibfnamefont {P.~A.}\ \bibnamefont {Morris}}, \bibinfo
  {author} {\bibfnamefont {D.~S.}\ \bibnamefont {Tasca}}, \bibinfo {author}
  {\bibfnamefont {L.}~\bibnamefont {Mertens}}, \bibinfo {author} {\bibfnamefont
  {M.~G.}\ \bibnamefont {Tanner}}, \bibinfo {author} {\bibfnamefont {R.~A.}\
  \bibnamefont {Kirkwood}}, \bibinfo {author} {\bibfnamefont {A.}~\bibnamefont
  {Ruggeri}}, \bibinfo {author} {\bibfnamefont {A.}~\bibnamefont {Tosi}},
  \bibinfo {author} {\bibfnamefont {R.~W.}\ \bibnamefont {Boyd}}, \bibinfo
  {author} {\bibfnamefont {G.~S.}\ \bibnamefont {Buller}}, \bibinfo {author}
  {\bibfnamefont {R.~H.}\ \bibnamefont {Hadfield}},\ and\ \bibinfo {author}
  {\bibfnamefont {M.~J.}\ \bibnamefont {Padgett}},\ }\bibfield  {title}
  {\bibinfo {title} {Photon-sparse microscopy: visible light imaging using
  infrared illumination},\ }\href {https://doi.org/10.1364/OPTICA.2.001049}
  {\bibfield  {journal} {\bibinfo  {journal} {Optica}\ }\textbf {\bibinfo
  {volume} {2}},\ \bibinfo {pages} {1049} (\bibinfo {year} {2015})}\BibitemShut
  {NoStop}%
\bibitem [{\citenamefont {Ryan}\ \emph {et~al.}(2024)\citenamefont {Ryan},
  \citenamefont {Meier}, \citenamefont {Seitz}, \citenamefont {Hanson},
  \citenamefont {Morales}, \citenamefont {Palmer}, \citenamefont {Hanson},
  \citenamefont {Goodwin}, \citenamefont {Newell}, \citenamefont {Holmes},
  \citenamefont {Thompson},\ and\ \citenamefont {Werner}}]{Ryan2024}%
  \BibitemOpen
  \bibfield  {author} {\bibinfo {author} {\bibfnamefont {D.~P.}\ \bibnamefont
  {Ryan}}, \bibinfo {author} {\bibfnamefont {K.}~\bibnamefont {Meier}},
  \bibinfo {author} {\bibfnamefont {K.}~\bibnamefont {Seitz}}, \bibinfo
  {author} {\bibfnamefont {D.}~\bibnamefont {Hanson}}, \bibinfo {author}
  {\bibfnamefont {D.}~\bibnamefont {Morales}}, \bibinfo {author} {\bibfnamefont
  {D.~M.}\ \bibnamefont {Palmer}}, \bibinfo {author} {\bibfnamefont
  {B.}~\bibnamefont {Hanson}}, \bibinfo {author} {\bibfnamefont {P.~M.}\
  \bibnamefont {Goodwin}}, \bibinfo {author} {\bibfnamefont {R.}~\bibnamefont
  {Newell}}, \bibinfo {author} {\bibfnamefont {R.~M.}\ \bibnamefont {Holmes}},
  \bibinfo {author} {\bibfnamefont {D.}~\bibnamefont {Thompson}},\ and\
  \bibinfo {author} {\bibfnamefont {J.}~\bibnamefont {Werner}},\ }\bibfield
  {title} {\bibinfo {title} {Infrared quantum ghost imaging of living and
  undisturbed plants},\ }\href {https://doi.org/10.1364/OPTICA.527982}
  {\bibfield  {journal} {\bibinfo  {journal} {Optica}\ }\textbf {\bibinfo
  {volume} {11}},\ \bibinfo {pages} {1261} (\bibinfo {year}
  {2024})}\BibitemShut {NoStop}%
\bibitem [{\citenamefont {Nasr}\ \emph {et~al.}(2003)\citenamefont {Nasr},
  \citenamefont {Saleh}, \citenamefont {Sergienko},\ and\ \citenamefont
  {Teich}}]{Nasr2003}%
  \BibitemOpen
  \bibfield  {author} {\bibinfo {author} {\bibfnamefont {M.~B.}\ \bibnamefont
  {Nasr}}, \bibinfo {author} {\bibfnamefont {B.~E.~A.}\ \bibnamefont {Saleh}},
  \bibinfo {author} {\bibfnamefont {A.~V.}\ \bibnamefont {Sergienko}},\ and\
  \bibinfo {author} {\bibfnamefont {M.~C.}\ \bibnamefont {Teich}},\ }\bibfield
  {title} {\bibinfo {title} {Demonstration of dispersion-canceled
  quantum-optical coherence tomography},\ }\href
  {https://doi.org/10.1103/PhysRevLett.91.083601} {\bibfield  {journal}
  {\bibinfo  {journal} {Phys. Rev. Lett.}\ }\textbf {\bibinfo {volume} {91}},\
  \bibinfo {pages} {083601} (\bibinfo {year} {2003})}\BibitemShut {NoStop}%
\bibitem [{\citenamefont {Yepiz-Graciano}\ \emph {et~al.}(2022)\citenamefont
  {Yepiz-Graciano}, \citenamefont {Ibarra-Borja}, \citenamefont
  {Ram\'{\i}rez~Alarc\'on}, \citenamefont {Guti\'errez-Torres}, \citenamefont
  {Cruz-Ram\'{\i}rez}, \citenamefont {Lopez-Mago},\ and\ \citenamefont
  {U'Ren}}]{Yepiz2022}%
  \BibitemOpen
  \bibfield  {author} {\bibinfo {author} {\bibfnamefont {P.}~\bibnamefont
  {Yepiz-Graciano}}, \bibinfo {author} {\bibfnamefont {Z.}~\bibnamefont
  {Ibarra-Borja}}, \bibinfo {author} {\bibfnamefont {R.}~\bibnamefont
  {Ram\'{\i}rez~Alarc\'on}}, \bibinfo {author} {\bibfnamefont {G.}~\bibnamefont
  {Guti\'errez-Torres}}, \bibinfo {author} {\bibfnamefont {H.}~\bibnamefont
  {Cruz-Ram\'{\i}rez}}, \bibinfo {author} {\bibfnamefont {D.}~\bibnamefont
  {Lopez-Mago}},\ and\ \bibinfo {author} {\bibfnamefont {A.~B.}\ \bibnamefont
  {U'Ren}},\ }\bibfield  {title} {\bibinfo {title} {Quantum optical coherence
  microscopy for bioimaging applications},\ }\href
  {https://doi.org/10.1103/PhysRevApplied.18.034060} {\bibfield  {journal}
  {\bibinfo  {journal} {Phys. Rev. Applied}\ }\textbf {\bibinfo {volume}
  {18}},\ \bibinfo {pages} {034060} (\bibinfo {year} {2022})}\BibitemShut
  {NoStop}%
\bibitem [{\citenamefont {{Zhang}}\ \emph {et~al.}(2023)\citenamefont
  {{Zhang}}, \citenamefont {{England}},\ and\ \citenamefont
  {{Sussman}}}]{Zhang20222}%
  \BibitemOpen
  \bibfield  {author} {\bibinfo {author} {\bibfnamefont {Y.}~\bibnamefont
  {{Zhang}}}, \bibinfo {author} {\bibfnamefont {D.}~\bibnamefont {{England}}},\
  and\ \bibinfo {author} {\bibfnamefont {B.}~\bibnamefont {{Sussman}}},\
  }\bibfield  {title} {\bibinfo {title} {Snapshot hyperspectral imaging with
  quantum correlated photons},\ }\href@noop {} {\bibfield  {journal} {\bibinfo
  {journal} {Opt. Express}\ }\textbf {\bibinfo {volume} {31}},\ \bibinfo
  {pages} {2282} (\bibinfo {year} {2023})}\BibitemShut {NoStop}%
\bibitem [{\citenamefont {Zhang}\ \emph {et~al.}(2022)\citenamefont {Zhang},
  \citenamefont {Orth}, \citenamefont {England},\ and\ \citenamefont
  {Sussman}}]{Zhang2022}%
  \BibitemOpen
  \bibfield  {author} {\bibinfo {author} {\bibfnamefont {Y.}~\bibnamefont
  {Zhang}}, \bibinfo {author} {\bibfnamefont {A.}~\bibnamefont {Orth}},
  \bibinfo {author} {\bibfnamefont {D.}~\bibnamefont {England}},\ and\ \bibinfo
  {author} {\bibfnamefont {B.}~\bibnamefont {Sussman}},\ }\bibfield  {title}
  {\bibinfo {title} {Ray tracing with quantum correlated photons to image a
  three-dimensional scene},\ }\href
  {https://doi.org/10.1103/PhysRevA.105.L011701} {\bibfield  {journal}
  {\bibinfo  {journal} {Phys. Rev. A}\ }\textbf {\bibinfo {volume} {105}},\
  \bibinfo {pages} {L011701} (\bibinfo {year} {2022})}\BibitemShut {NoStop}%
\bibitem [{\citenamefont {Zhang}\ \emph {et~al.}(2024)\citenamefont {Zhang},
  \citenamefont {England}, \citenamefont {Orth}, \citenamefont {Karimi},\ and\
  \citenamefont {Sussman}}]{Zhang2024}%
  \BibitemOpen
  \bibfield  {author} {\bibinfo {author} {\bibfnamefont {Y.}~\bibnamefont
  {Zhang}}, \bibinfo {author} {\bibfnamefont {D.}~\bibnamefont {England}},
  \bibinfo {author} {\bibfnamefont {A.}~\bibnamefont {Orth}}, \bibinfo {author}
  {\bibfnamefont {E.}~\bibnamefont {Karimi}},\ and\ \bibinfo {author}
  {\bibfnamefont {B.}~\bibnamefont {Sussman}},\ }\bibfield  {title} {\bibinfo
  {title} {Quantum light-field microscopy for volumetric imaging with extreme
  depth of field},\ }\href {https://doi.org/10.1103/PhysRevApplied.21.024029}
  {\bibfield  {journal} {\bibinfo  {journal} {Phys. Rev. Appl.}\ }\textbf
  {\bibinfo {volume} {21}},\ \bibinfo {pages} {024029} (\bibinfo {year}
  {2024})}\BibitemShut {NoStop}%
\bibitem [{\citenamefont {Ono}\ \emph {et~al.}(2013)\citenamefont {Ono},
  \citenamefont {Okamoto},\ and\ \citenamefont {Takeuchi}}]{Ono2013}%
  \BibitemOpen
  \bibfield  {author} {\bibinfo {author} {\bibfnamefont {T.}~\bibnamefont
  {Ono}}, \bibinfo {author} {\bibfnamefont {R.}~\bibnamefont {Okamoto}},\ and\
  \bibinfo {author} {\bibfnamefont {S.}~\bibnamefont {Takeuchi}},\ }\bibfield
  {title} {\bibinfo {title} {An entanglement-enhanced microscope},\ }\href
  {https://doi.org/10.1038/ncomms3426} {\bibfield  {journal} {\bibinfo
  {journal} {Nature Communications}\ }\textbf {\bibinfo {volume} {4}},\
  \bibinfo {pages} {2426} (\bibinfo {year} {2013})}\BibitemShut {NoStop}%
\bibitem [{\citenamefont {Israel}\ \emph {et~al.}(2014)\citenamefont {Israel},
  \citenamefont {Rosen},\ and\ \citenamefont {Silberberg}}]{Israe2014}%
  \BibitemOpen
  \bibfield  {author} {\bibinfo {author} {\bibfnamefont {Y.}~\bibnamefont
  {Israel}}, \bibinfo {author} {\bibfnamefont {S.}~\bibnamefont {Rosen}},\ and\
  \bibinfo {author} {\bibfnamefont {Y.}~\bibnamefont {Silberberg}},\ }\bibfield
   {title} {\bibinfo {title} {Supersensitive polarization microscopy using noon
  states of light},\ }\href {https://doi.org/10.1103/PhysRevLett.112.103604}
  {\bibfield  {journal} {\bibinfo  {journal} {Phys. Rev. Lett.}\ }\textbf
  {\bibinfo {volume} {112}},\ \bibinfo {pages} {103604} (\bibinfo {year}
  {2014})}\BibitemShut {NoStop}%
\bibitem [{\citenamefont {Black}\ \emph {et~al.}(2023)\citenamefont {Black},
  \citenamefont {Nguyen}, \citenamefont {Braverman}, \citenamefont {Crampton},
  \citenamefont {Evans},\ and\ \citenamefont {Boyd}}]{Black2023}%
  \BibitemOpen
  \bibfield  {author} {\bibinfo {author} {\bibfnamefont {A.~N.}\ \bibnamefont
  {Black}}, \bibinfo {author} {\bibfnamefont {L.~D.}\ \bibnamefont {Nguyen}},
  \bibinfo {author} {\bibfnamefont {B.}~\bibnamefont {Braverman}}, \bibinfo
  {author} {\bibfnamefont {K.~T.}\ \bibnamefont {Crampton}}, \bibinfo {author}
  {\bibfnamefont {J.~E.}\ \bibnamefont {Evans}},\ and\ \bibinfo {author}
  {\bibfnamefont {R.~W.}\ \bibnamefont {Boyd}},\ }\bibfield  {title} {\bibinfo
  {title} {Quantum-enhanced phase imaging without coincidence counting},\
  }\href {https://doi.org/10.1364/OPTICA.482926} {\bibfield  {journal}
  {\bibinfo  {journal} {Optica}\ }\textbf {\bibinfo {volume} {10}},\ \bibinfo
  {pages} {952} (\bibinfo {year} {2023})}\BibitemShut {NoStop}%
\bibitem [{\citenamefont {Defienne}\ \emph
  {et~al.}(2021{\natexlab{b}})\citenamefont {Defienne}, \citenamefont
  {Ndagano}, \citenamefont {Lyons},\ and\ \citenamefont
  {Faccio}}]{Defienne2021}%
  \BibitemOpen
  \bibfield  {author} {\bibinfo {author} {\bibfnamefont {H.}~\bibnamefont
  {Defienne}}, \bibinfo {author} {\bibfnamefont {B.}~\bibnamefont {Ndagano}},
  \bibinfo {author} {\bibfnamefont {A.}~\bibnamefont {Lyons}},\ and\ \bibinfo
  {author} {\bibfnamefont {D.}~\bibnamefont {Faccio}},\ }\bibfield  {title}
  {\bibinfo {title} {Polarization entanglement-enabled quantum holography},\
  }\href {https://doi.org/10.1038/s41567-020-01156-1} {\bibfield  {journal}
  {\bibinfo  {journal} {Nature Physics}\ }\textbf {\bibinfo {volume} {17}},\
  \bibinfo {pages} {591} (\bibinfo {year} {2021}{\natexlab{b}})}\BibitemShut
  {NoStop}%
\bibitem [{\citenamefont {Thekkadath}\ \emph {et~al.}(2023)\citenamefont
  {Thekkadath}, \citenamefont {England}, \citenamefont {Bouchard},
  \citenamefont {Zhang}, \citenamefont {Kim},\ and\ \citenamefont
  {Sussman}}]{Guillaume2023}%
  \BibitemOpen
  \bibfield  {author} {\bibinfo {author} {\bibfnamefont {G.}~\bibnamefont
  {Thekkadath}}, \bibinfo {author} {\bibfnamefont {D.}~\bibnamefont {England}},
  \bibinfo {author} {\bibfnamefont {F.}~\bibnamefont {Bouchard}}, \bibinfo
  {author} {\bibfnamefont {Y.}~\bibnamefont {Zhang}}, \bibinfo {author}
  {\bibfnamefont {M.}~\bibnamefont {Kim}},\ and\ \bibinfo {author}
  {\bibfnamefont {B.}~\bibnamefont {Sussman}},\ }\bibfield  {title} {\bibinfo
  {title} {Intensity interferometry for holography with quantum and classical
  light},\ }\href {https://doi.org/10.1126/sciadv.adh1439} {\bibfield
  {journal} {\bibinfo  {journal} {Science Advances}\ }\textbf {\bibinfo
  {volume} {9}},\ \bibinfo {pages} {eadh1439} (\bibinfo {year}
  {2023})}\BibitemShut {NoStop}%
\bibitem [{\citenamefont {Abouraddy}\ \emph {et~al.}(2004)\citenamefont
  {Abouraddy}, \citenamefont {Stone}, \citenamefont {Sergienko}, \citenamefont
  {Saleh},\ and\ \citenamefont {Teich}}]{Abouraddy2004}%
  \BibitemOpen
  \bibfield  {author} {\bibinfo {author} {\bibfnamefont {A.~F.}\ \bibnamefont
  {Abouraddy}}, \bibinfo {author} {\bibfnamefont {P.~R.}\ \bibnamefont
  {Stone}}, \bibinfo {author} {\bibfnamefont {A.~V.}\ \bibnamefont
  {Sergienko}}, \bibinfo {author} {\bibfnamefont {B.~E.~A.}\ \bibnamefont
  {Saleh}},\ and\ \bibinfo {author} {\bibfnamefont {M.~C.}\ \bibnamefont
  {Teich}},\ }\bibfield  {title} {\bibinfo {title} {Entangled-photon imaging of
  a pure phase object},\ }\href {https://doi.org/10.1103/PhysRevLett.93.213903}
  {\bibfield  {journal} {\bibinfo  {journal} {Phys. Rev. Lett.}\ }\textbf
  {\bibinfo {volume} {93}},\ \bibinfo {pages} {213903} (\bibinfo {year}
  {2004})}\BibitemShut {NoStop}%
\bibitem [{\citenamefont {Sephton}\ \emph {et~al.}(2023)\citenamefont
  {Sephton}, \citenamefont {Nape}, \citenamefont {Moodley}, \citenamefont
  {Francis},\ and\ \citenamefont {Forbes}}]{Sephton2023}%
  \BibitemOpen
  \bibfield  {author} {\bibinfo {author} {\bibfnamefont {B.}~\bibnamefont
  {Sephton}}, \bibinfo {author} {\bibfnamefont {I.}~\bibnamefont {Nape}},
  \bibinfo {author} {\bibfnamefont {C.}~\bibnamefont {Moodley}}, \bibinfo
  {author} {\bibfnamefont {J.}~\bibnamefont {Francis}},\ and\ \bibinfo {author}
  {\bibfnamefont {A.}~\bibnamefont {Forbes}},\ }\bibfield  {title} {\bibinfo
  {title} {Revealing the embedded phase in single-pixel quantum ghost
  imaging},\ }\href {https://doi.org/10.1364/OPTICA.472980} {\bibfield
  {journal} {\bibinfo  {journal} {Optica}\ }\textbf {\bibinfo {volume} {10}},\
  \bibinfo {pages} {286} (\bibinfo {year} {2023})}\BibitemShut {NoStop}%
\bibitem [{\citenamefont {Lu}\ \emph {et~al.}(2015)\citenamefont {Lu},
  \citenamefont {Reichert}, \citenamefont {Sun},\ and\ \citenamefont
  {Fleischer}}]{Chien2015}%
  \BibitemOpen
  \bibfield  {author} {\bibinfo {author} {\bibfnamefont {C.-H.}\ \bibnamefont
  {Lu}}, \bibinfo {author} {\bibfnamefont {M.}~\bibnamefont {Reichert}},
  \bibinfo {author} {\bibfnamefont {X.}~\bibnamefont {Sun}},\ and\ \bibinfo
  {author} {\bibfnamefont {J.~W.}\ \bibnamefont {Fleischer}},\ }\href@noop {}
  {\bibinfo {title} {Quantum phase imaging using spatial entanglement}}
  (\bibinfo {year} {2015}),\ \Eprint {https://arxiv.org/abs/1509.01227}
  {arXiv:1509.01227 [physics.optics]} \BibitemShut {NoStop}%
\bibitem [{\citenamefont {Ortolano}\ \emph {et~al.}(2023)\citenamefont
  {Ortolano}, \citenamefont {Paniate}, \citenamefont {Boucher}, \citenamefont
  {Napoli}, \citenamefont {Soman}, \citenamefont {Pereira}, \citenamefont
  {Ruo-Berchera},\ and\ \citenamefont {Genovese}}]{Ortolano2023}%
  \BibitemOpen
  \bibfield  {author} {\bibinfo {author} {\bibfnamefont {G.}~\bibnamefont
  {Ortolano}}, \bibinfo {author} {\bibfnamefont {A.}~\bibnamefont {Paniate}},
  \bibinfo {author} {\bibfnamefont {P.}~\bibnamefont {Boucher}}, \bibinfo
  {author} {\bibfnamefont {C.}~\bibnamefont {Napoli}}, \bibinfo {author}
  {\bibfnamefont {S.}~\bibnamefont {Soman}}, \bibinfo {author} {\bibfnamefont
  {S.~F.}\ \bibnamefont {Pereira}}, \bibinfo {author} {\bibfnamefont
  {I.}~\bibnamefont {Ruo-Berchera}},\ and\ \bibinfo {author} {\bibfnamefont
  {M.}~\bibnamefont {Genovese}},\ }\bibfield  {title} {\bibinfo {title}
  {Quantum enhanced non-interferometric quantitative phase imaging},\ }\href
  {https://doi.org/10.1038/s41377-023-01215-1} {\bibfield  {journal} {\bibinfo
  {journal} {Light: Science {\&} Applications}\ }\textbf {\bibinfo {volume}
  {12}},\ \bibinfo {pages} {171} (\bibinfo {year} {2023})}\BibitemShut
  {NoStop}%
\bibitem [{\citenamefont {Aidukas}\ \emph {et~al.}(2019)\citenamefont
  {Aidukas}, \citenamefont {Konda}, \citenamefont {Harvey}, \citenamefont
  {Padgett},\ and\ \citenamefont {Moreau}}]{Aidukas2019}%
  \BibitemOpen
  \bibfield  {author} {\bibinfo {author} {\bibfnamefont {T.}~\bibnamefont
  {Aidukas}}, \bibinfo {author} {\bibfnamefont {P.~C.}\ \bibnamefont {Konda}},
  \bibinfo {author} {\bibfnamefont {A.~R.}\ \bibnamefont {Harvey}}, \bibinfo
  {author} {\bibfnamefont {M.~J.}\ \bibnamefont {Padgett}},\ and\ \bibinfo
  {author} {\bibfnamefont {P.-A.}\ \bibnamefont {Moreau}},\ }\bibfield  {title}
  {\bibinfo {title} {Phase and amplitude imaging with quantum correlations
  through fourier ptychography},\ }\href
  {https://doi.org/10.1038/s41598-019-46273-x} {\bibfield  {journal} {\bibinfo
  {journal} {Scientific Reports}\ }\textbf {\bibinfo {volume} {9}},\ \bibinfo
  {pages} {10445} (\bibinfo {year} {2019})}\BibitemShut {NoStop}%
\bibitem [{\citenamefont {Hodgson}\ \emph {et~al.}(2023)\citenamefont
  {Hodgson}, \citenamefont {Zhang}, \citenamefont {England},\ and\
  \citenamefont {Sussman}}]{Hodgson2022}%
  \BibitemOpen
  \bibfield  {author} {\bibinfo {author} {\bibfnamefont {H.}~\bibnamefont
  {Hodgson}}, \bibinfo {author} {\bibfnamefont {Y.}~\bibnamefont {Zhang}},
  \bibinfo {author} {\bibfnamefont {D.}~\bibnamefont {England}},\ and\ \bibinfo
  {author} {\bibfnamefont {B.}~\bibnamefont {Sussman}},\ }\bibfield  {title}
  {\bibinfo {title} {Reconfigurable phase contrast microscopy with correlated
  photon pairs},\ }\href {https://doi.org/10.1063/5.0133980} {\bibfield
  {journal} {\bibinfo  {journal} {Applied Physics Letters}\ }\textbf {\bibinfo
  {volume} {122}},\ \bibinfo {pages} {034001} (\bibinfo {year}
  {2023})}\BibitemShut {NoStop}%
\bibitem [{\citenamefont {Zheng}\ \emph {et~al.}(2024)\citenamefont {Zheng},
  \citenamefont {Liu}, \citenamefont {Miao}, \citenamefont {Cui}, \citenamefont
  {Yang}, \citenamefont {Xu}, \citenamefont {Xu}, \citenamefont {Li},\ and\
  \citenamefont {Guo}}]{Zheng2024}%
  \BibitemOpen
  \bibfield  {author} {\bibinfo {author} {\bibfnamefont {Y.}~\bibnamefont
  {Zheng}}, \bibinfo {author} {\bibfnamefont {Z.-D.}\ \bibnamefont {Liu}},
  \bibinfo {author} {\bibfnamefont {R.-H.}\ \bibnamefont {Miao}}, \bibinfo
  {author} {\bibfnamefont {J.-M.}\ \bibnamefont {Cui}}, \bibinfo {author}
  {\bibfnamefont {M.}~\bibnamefont {Yang}}, \bibinfo {author} {\bibfnamefont
  {X.-Y.}\ \bibnamefont {Xu}}, \bibinfo {author} {\bibfnamefont {J.-S.}\
  \bibnamefont {Xu}}, \bibinfo {author} {\bibfnamefont {C.-F.}\ \bibnamefont
  {Li}},\ and\ \bibinfo {author} {\bibfnamefont {G.-C.}\ \bibnamefont {Guo}},\
  }\bibfield  {title} {\bibinfo {title} {Characterizing biphoton spatial wave
  function dynamics with quantum wavefront sensing},\ }\href
  {https://doi.org/10.1103/PhysRevLett.133.033602} {\bibfield  {journal}
  {\bibinfo  {journal} {Phys. Rev. Lett.}\ }\textbf {\bibinfo {volume} {133}},\
  \bibinfo {pages} {033602} (\bibinfo {year} {2024})}\BibitemShut {NoStop}%
\bibitem [{\citenamefont {Dehghan}\ \emph {et~al.}(2024)\citenamefont
  {Dehghan}, \citenamefont {D'Errico}, \citenamefont {Colandrea},\ and\
  \citenamefont {Karimi}}]{Dehghan2024}%
  \BibitemOpen
  \bibfield  {author} {\bibinfo {author} {\bibfnamefont {N.}~\bibnamefont
  {Dehghan}}, \bibinfo {author} {\bibfnamefont {A.}~\bibnamefont {D'Errico}},
  \bibinfo {author} {\bibfnamefont {F.~D.}\ \bibnamefont {Colandrea}},\ and\
  \bibinfo {author} {\bibfnamefont {E.}~\bibnamefont {Karimi}},\ }\bibfield
  {title} {\bibinfo {title} {Biphoton state reconstruction via phase retrieval
  methods},\ }\href {https://doi.org/10.1364/OPTICA.527661} {\bibfield
  {journal} {\bibinfo  {journal} {Optica}\ }\textbf {\bibinfo {volume} {11}},\
  \bibinfo {pages} {1115} (\bibinfo {year} {2024})}\BibitemShut {NoStop}%
\bibitem [{\citenamefont {Law}\ and\ \citenamefont {Eberly}(2004)}]{Law2004}%
  \BibitemOpen
  \bibfield  {author} {\bibinfo {author} {\bibfnamefont {C.~K.}\ \bibnamefont
  {Law}}\ and\ \bibinfo {author} {\bibfnamefont {J.~H.}\ \bibnamefont
  {Eberly}},\ }\bibfield  {title} {\bibinfo {title} {Analysis and
  interpretation of high transverse entanglement in optical parametric down
  conversion},\ }\href {https://doi.org/10.1103/PhysRevLett.92.127903}
  {\bibfield  {journal} {\bibinfo  {journal} {Phys. Rev. Lett.}\ }\textbf
  {\bibinfo {volume} {92}},\ \bibinfo {pages} {127903} (\bibinfo {year}
  {2004})}\BibitemShut {NoStop}%
\bibitem [{\citenamefont {Chan}\ \emph {et~al.}(2007)\citenamefont {Chan},
  \citenamefont {Torres},\ and\ \citenamefont {Eberly}}]{Chan2007}%
  \BibitemOpen
  \bibfield  {author} {\bibinfo {author} {\bibfnamefont {K.~W.}\ \bibnamefont
  {Chan}}, \bibinfo {author} {\bibfnamefont {J.~P.}\ \bibnamefont {Torres}},\
  and\ \bibinfo {author} {\bibfnamefont {J.~H.}\ \bibnamefont {Eberly}},\
  }\bibfield  {title} {\bibinfo {title} {Transverse entanglement migration in
  hilbert space},\ }\href {https://doi.org/10.1103/PhysRevA.75.050101}
  {\bibfield  {journal} {\bibinfo  {journal} {Phys. Rev. A}\ }\textbf {\bibinfo
  {volume} {75}},\ \bibinfo {pages} {050101} (\bibinfo {year}
  {2007})}\BibitemShut {NoStop}%
\bibitem [{\citenamefont {Schneeloch}\ and\ \citenamefont
  {Howell}(2016)}]{Schneeloch2016}%
  \BibitemOpen
  \bibfield  {author} {\bibinfo {author} {\bibfnamefont {J.}~\bibnamefont
  {Schneeloch}}\ and\ \bibinfo {author} {\bibfnamefont {J.~C.}\ \bibnamefont
  {Howell}},\ }\bibfield  {title} {\bibinfo {title} {Introduction to the
  transverse spatial correlations in spontaneous parametric down-conversion
  through the biphoton birth zone},\ }\href
  {https://doi.org/10.1088/2040-8978/18/5/053501} {\bibfield  {journal}
  {\bibinfo  {journal} {Journal of Optics}\ }\textbf {\bibinfo {volume} {18}},\
  \bibinfo {pages} {053501} (\bibinfo {year} {2016})}\BibitemShut {NoStop}%
\bibitem [{\citenamefont {Nomerotski}(2019)}]{Nomerotski2019}%
  \BibitemOpen
  \bibfield  {author} {\bibinfo {author} {\bibfnamefont {A.}~\bibnamefont
  {Nomerotski}},\ }\bibfield  {title} {\bibinfo {title} {Imaging and time
  stamping of photons with nanosecond resolution in timepix based optical
  cameras},\ }\href
  {https://doi.org/https://doi.org/10.1016/j.nima.2019.05.034} {\bibfield
  {journal} {\bibinfo  {journal} {Nuclear Instruments and Methods in Physics
  Research Section A: Accelerators, Spectrometers, Detectors and Associated
  Equipment}\ }\textbf {\bibinfo {volume} {937}},\ \bibinfo {pages} {26 }
  (\bibinfo {year} {2019})}\BibitemShut {NoStop}%
\bibitem [{ASI({\natexlab{a}})}]{ASI}%
  \BibitemOpen
  \href {https://kt.cern/technologies/timepix3} {\bibinfo {title}
  {https://kt.cern/technologies/timepix3}} ({\natexlab{a}})\BibitemShut
  {NoStop}%
\bibitem [{\citenamefont {Frankot}\ and\ \citenamefont
  {Chellappa}(1988)}]{Frankot1988}%
  \BibitemOpen
  \bibfield  {author} {\bibinfo {author} {\bibfnamefont {R.}~\bibnamefont
  {Frankot}}\ and\ \bibinfo {author} {\bibfnamefont {R.}~\bibnamefont
  {Chellappa}},\ }\bibfield  {title} {\bibinfo {title} {A method for enforcing
  integrability in shape from shading algorithms},\ }\href
  {https://doi.org/10.1109/34.3909} {\bibfield  {journal} {\bibinfo  {journal}
  {IEEE Transactions on Pattern Analysis and Machine Intelligence}\ }\textbf
  {\bibinfo {volume} {10}},\ \bibinfo {pages} {439} (\bibinfo {year}
  {1988})}\BibitemShut {NoStop}%
\bibitem [{Ben()}]{Benchmark}%
  \BibitemOpen
  \href {https://benchmarktech.com/quantitativephasemicroscop/} {\bibinfo
  {title} {https://benchmarktech.com/quantitativephasemicroscop/}}\BibitemShut
  {NoStop}%
\bibitem [{\citenamefont {Gul}\ \emph {et~al.}(2021)\citenamefont {Gul},
  \citenamefont {Ashraf}, \citenamefont {Khan}, \citenamefont {Nisar},\ and\
  \citenamefont {Ahmad}}]{Gul2021}%
  \BibitemOpen
  \bibfield  {author} {\bibinfo {author} {\bibfnamefont {B.}~\bibnamefont
  {Gul}}, \bibinfo {author} {\bibfnamefont {S.}~\bibnamefont {Ashraf}},
  \bibinfo {author} {\bibfnamefont {S.}~\bibnamefont {Khan}}, \bibinfo {author}
  {\bibfnamefont {H.}~\bibnamefont {Nisar}},\ and\ \bibinfo {author}
  {\bibfnamefont {I.}~\bibnamefont {Ahmad}},\ }\bibfield  {title} {\bibinfo
  {title} {Cell refractive index: Models, insights, applications and future
  perspectives},\ }\href
  {https://doi.org/https://doi.org/10.1016/j.pdpdt.2020.102096} {\bibfield
  {journal} {\bibinfo  {journal} {Photodiagnosis and Photodynamic Therapy}\
  }\textbf {\bibinfo {volume} {33}},\ \bibinfo {pages} {102096} (\bibinfo
  {year} {2021})}\BibitemShut {NoStop}%
\bibitem [{\citenamefont {England}\ \emph {et~al.}(2019)\citenamefont
  {England}, \citenamefont {Balaji},\ and\ \citenamefont
  {Sussman}}]{England2019}%
  \BibitemOpen
  \bibfield  {author} {\bibinfo {author} {\bibfnamefont {D.~G.}\ \bibnamefont
  {England}}, \bibinfo {author} {\bibfnamefont {B.}~\bibnamefont {Balaji}},\
  and\ \bibinfo {author} {\bibfnamefont {B.~J.}\ \bibnamefont {Sussman}},\
  }\bibfield  {title} {\bibinfo {title} {Quantum-enhanced standoff detection
  using correlated photon pairs},\ }\href
  {https://doi.org/10.1103/PhysRevA.99.023828} {\bibfield  {journal} {\bibinfo
  {journal} {Phys. Rev. A}\ }\textbf {\bibinfo {volume} {99}},\ \bibinfo
  {pages} {023828} (\bibinfo {year} {2019})}\BibitemShut {NoStop}%
\bibitem [{\citenamefont {Vidyapin}\ \emph {et~al.}(2023)\citenamefont
  {Vidyapin}, \citenamefont {Zhang}, \citenamefont {England},\ and\
  \citenamefont {Sussman}}]{Vidyapin2022}%
  \BibitemOpen
  \bibfield  {author} {\bibinfo {author} {\bibfnamefont {V.}~\bibnamefont
  {Vidyapin}}, \bibinfo {author} {\bibfnamefont {Y.}~\bibnamefont {Zhang}},
  \bibinfo {author} {\bibfnamefont {D.}~\bibnamefont {England}},\ and\ \bibinfo
  {author} {\bibfnamefont {B.}~\bibnamefont {Sussman}},\ }\bibfield  {title}
  {\bibinfo {title} {Characterisation of a single photon event camera for
  quantum imaging},\ }\href {https://doi.org/10.1038/s41598-023-27842-7}
  {\bibfield  {journal} {\bibinfo  {journal} {Scientific Reports}\ }\textbf
  {\bibinfo {volume} {13}},\ \bibinfo {pages} {1009} (\bibinfo {year}
  {2023})}\BibitemShut {NoStop}%
\bibitem [{\citenamefont {Wollman}\ \emph {et~al.}(2019)\citenamefont
  {Wollman}, \citenamefont {Verma}, \citenamefont {Lita}, \citenamefont {Farr},
  \citenamefont {Shaw}, \citenamefont {Mirin},\ and\ \citenamefont
  {Nam}}]{Wollman2019}%
  \BibitemOpen
  \bibfield  {author} {\bibinfo {author} {\bibfnamefont {E.~E.}\ \bibnamefont
  {Wollman}}, \bibinfo {author} {\bibfnamefont {V.~B.}\ \bibnamefont {Verma}},
  \bibinfo {author} {\bibfnamefont {A.~E.}\ \bibnamefont {Lita}}, \bibinfo
  {author} {\bibfnamefont {W.~H.}\ \bibnamefont {Farr}}, \bibinfo {author}
  {\bibfnamefont {M.~D.}\ \bibnamefont {Shaw}}, \bibinfo {author}
  {\bibfnamefont {R.~P.}\ \bibnamefont {Mirin}},\ and\ \bibinfo {author}
  {\bibfnamefont {S.~W.}\ \bibnamefont {Nam}},\ }\bibfield  {title} {\bibinfo
  {title} {Kilopixel array of superconducting nanowire single-photon
  detectors},\ }\href {https://doi.org/10.1364/OE.27.035279} {\bibfield
  {journal} {\bibinfo  {journal} {Opt. Express}\ }\textbf {\bibinfo {volume}
  {27}},\ \bibinfo {pages} {35279} (\bibinfo {year} {2019})}\BibitemShut
  {NoStop}%
\bibitem [{\citenamefont {Oripov}\ \emph {et~al.}(2023)\citenamefont {Oripov},
  \citenamefont {Rampini}, \citenamefont {Allmaras}, \citenamefont {Shaw},
  \citenamefont {Nam}, \citenamefont {Korzh},\ and\ \citenamefont
  {McCaughan}}]{Oripov2023}%
  \BibitemOpen
  \bibfield  {author} {\bibinfo {author} {\bibfnamefont {B.~G.}\ \bibnamefont
  {Oripov}}, \bibinfo {author} {\bibfnamefont {D.~S.}\ \bibnamefont {Rampini}},
  \bibinfo {author} {\bibfnamefont {J.}~\bibnamefont {Allmaras}}, \bibinfo
  {author} {\bibfnamefont {M.~D.}\ \bibnamefont {Shaw}}, \bibinfo {author}
  {\bibfnamefont {S.~W.}\ \bibnamefont {Nam}}, \bibinfo {author} {\bibfnamefont
  {B.}~\bibnamefont {Korzh}},\ and\ \bibinfo {author} {\bibfnamefont {A.~N.}\
  \bibnamefont {McCaughan}},\ }\bibfield  {title} {\bibinfo {title} {A
  superconducting nanowire single-photon camera with 400,000 pixels},\ }\href
  {https://doi.org/10.1038/s41586-023-06550-2} {\bibfield  {journal} {\bibinfo
  {journal} {Nature}\ }\textbf {\bibinfo {volume} {622}},\ \bibinfo {pages}
  {730} (\bibinfo {year} {2023})}\BibitemShut {NoStop}%
\bibitem [{\citenamefont {Tyson}\ and\ \citenamefont
  {Frazier}(2022)}]{Tyson2022}%
  \BibitemOpen
  \bibfield  {author} {\bibinfo {author} {\bibfnamefont {R.}~\bibnamefont
  {Tyson}}\ and\ \bibinfo {author} {\bibfnamefont {B.}~\bibnamefont
  {Frazier}},\ }\href {https://books.google.ca/books?id=GhyTzgEACAAJ} {\emph
  {\bibinfo {title} {Principles of Adaptive Optics}}}\ (\bibinfo  {publisher}
  {CRC Press},\ \bibinfo {year} {2022})\BibitemShut {NoStop}%
\bibitem [{\citenamefont {Neal}\ \emph {et~al.}(2002)\citenamefont {Neal},
  \citenamefont {Copland},\ and\ \citenamefont {Neal}}]{Neal2002}%
  \BibitemOpen
  \bibfield  {author} {\bibinfo {author} {\bibfnamefont {D.~R.}\ \bibnamefont
  {Neal}}, \bibinfo {author} {\bibfnamefont {J.}~\bibnamefont {Copland}},\ and\
  \bibinfo {author} {\bibfnamefont {D.~A.}\ \bibnamefont {Neal}},\ }\bibfield
  {title} {\bibinfo {title} {Shack-hartmann wavefront sensor precision and
  accuracy},\ }in\ \href@noop {} {\emph {\bibinfo {booktitle} {Advanced
  Characterization Techniques for Optical, Semiconductor, and Data Storage
  Components}}},\ Vol.\ \bibinfo {volume} {4779}\ (\bibinfo {organization}
  {SPIE},\ \bibinfo {year} {2002})\ pp.\ \bibinfo {pages}
  {148--160}\BibitemShut {NoStop}%
\bibitem [{\citenamefont {Alexander}\ and\ \citenamefont
  {Ng}(1991)}]{Alexander1991}%
  \BibitemOpen
  \bibfield  {author} {\bibinfo {author} {\bibfnamefont {B.~F.}\ \bibnamefont
  {Alexander}}\ and\ \bibinfo {author} {\bibfnamefont {K.~C.}\ \bibnamefont
  {Ng}},\ }\bibfield  {title} {\bibinfo {title} {{Elimination of systematic
  error in subpixel accuracy centroid estimation [also Letter
  34(11)3347-3348(Nov1995)]}},\ }\href {https://doi.org/10.1117/12.55947}
  {\bibfield  {journal} {\bibinfo  {journal} {Optical Engineering}\ }\textbf
  {\bibinfo {volume} {30}},\ \bibinfo {pages} {1320 } (\bibinfo {year}
  {1991})}\BibitemShut {NoStop}%
\bibitem [{ASI({\natexlab{b}})}]{ASI1}%
  \BibitemOpen
  \href {https://www.amscins.com/product/cheetah-series} {\bibinfo {title}
  {https://www.amscins.com/product/cheetah-series}}
  ({\natexlab{b}})\BibitemShut {NoStop}%
\bibitem [{\citenamefont {Pfund}\ \emph {et~al.}(1998)\citenamefont {Pfund},
  \citenamefont {Lindlein},\ and\ \citenamefont {Schwider}}]{Pfund1998}%
  \BibitemOpen
  \bibfield  {author} {\bibinfo {author} {\bibfnamefont {J.}~\bibnamefont
  {Pfund}}, \bibinfo {author} {\bibfnamefont {N.}~\bibnamefont {Lindlein}},\
  and\ \bibinfo {author} {\bibfnamefont {J.}~\bibnamefont {Schwider}},\
  }\bibfield  {title} {\bibinfo {title} {Misalignment effects of the
  shack--hartmann sensor},\ }\href {https://doi.org/10.1364/AO.37.000022}
  {\bibfield  {journal} {\bibinfo  {journal} {Appl. Opt.}\ }\textbf {\bibinfo
  {volume} {37}},\ \bibinfo {pages} {22} (\bibinfo {year} {1998})}\BibitemShut
  {NoStop}%
\bibitem [{\citenamefont {Chernyshov}\ \emph {et~al.}(2005)\citenamefont
  {Chernyshov}, \citenamefont {Sterr}, \citenamefont {Riehle}, \citenamefont
  {Helmcke},\ and\ \citenamefont {Pfund}}]{Chernyshov2005}%
  \BibitemOpen
  \bibfield  {author} {\bibinfo {author} {\bibfnamefont {A.}~\bibnamefont
  {Chernyshov}}, \bibinfo {author} {\bibfnamefont {U.}~\bibnamefont {Sterr}},
  \bibinfo {author} {\bibfnamefont {F.}~\bibnamefont {Riehle}}, \bibinfo
  {author} {\bibfnamefont {J.}~\bibnamefont {Helmcke}},\ and\ \bibinfo {author}
  {\bibfnamefont {J.}~\bibnamefont {Pfund}},\ }\bibfield  {title} {\bibinfo
  {title} {Calibration of a shack--hartmann sensor for absolute measurements of
  wavefronts},\ }\href {https://doi.org/10.1364/AO.44.006419} {\bibfield
  {journal} {\bibinfo  {journal} {Appl. Opt.}\ }\textbf {\bibinfo {volume}
  {44}},\ \bibinfo {pages} {6419} (\bibinfo {year} {2005})}\BibitemShut
  {NoStop}%
\end{thebibliography}%
\end{document}